\documentclass[reprint, amsmath,amssymb, aps,]{revtex4-2}

\usepackage{graphicx}
\usepackage{bm}
\usepackage{chngcntr}

\begin{document}

\title{Self-Organization to the Edge of Ergodicity Breaking \\ in a Complex Adaptive System}

\author{Nixie Sapphira Lesmana$^{1, 2}$}
\author{Ling Feng$^{3,1}$}
\email{felney@gmail.com}
\author{Kan Chen$^{4,5}$}
\email{kan.chen@nus.edu.sg}
\author{Choy Heng Lai$^{1}$}
\affiliation{$^1$Department of Physics, National University of Singapore, Singapore}
\affiliation{$^2$NUS Cities, National University of Singapore, Singapore}
\affiliation{$^3$Institute of Advanced Intelligence and Computing (IAIC), Agency for Science, Technology and Research (A*STAR), Singapore}
\affiliation{$^4$Risk Management Institute, National University of Singapore, Singapore}
\affiliation{$^5$Department of Mathematics, National University of Singapore, Singapore}

\date{\today}

\begin{abstract}
Self-organized criticality is widely invoked for collective behavior, yet its role in objective-driven, heterogeneous adaptive systems is unclear. We introduce {\tt EvoSK}: agents learn on a Sherrington--Kirkpatrick landscape while the least fit are replaced. It self-organizes to the edge of ergodicity breaking, with scale-free avalanches ($\tau \approx -1.5$) and rewards beating any tuned non-evolutionary regime. Its cascade's branching ratio is the spectral radius of the learning dynamics' Jacobian, making critical branching and ergodicity breaking one marginal-stability condition fixing the exponent. The attraction to criticality follows from the selection--mutation balance: subcritical cascades decay too fast to dislodge frozen agents, supercritical cascades shield them from selection; only the critical power-law tail supplies the polynomial rate the balance requires.
\end{abstract}

\maketitle

\section{\label{sec:intro}Introduction}

Self-organized criticality (SOC) --- the idea that driven, dissipative systems tune themselves to a critical state --- has been proposed as a unifying mechanism for collective behavior across physics \cite{bak1987self,drossel1992self,clar1999self,zapperi1995self,pruessner2012self,cavagna2010scale}, biology \cite{bak1993punctuated,beggs2003neuronal}, and learning \cite{hopfield1982neural,feng2019optimal,qin2016collective,heins2024collective}. In canonical SOC models, scale-free avalanches arise from thresholds tied to a control parameter: critical density in forest fires \cite{drossel1992self,clar1999self}, slope in sandpiles \cite{manna1991critical,christensen1993sandpile}, or fitness in evolution \cite{bak1993punctuated,flyvbjerg1993mean}. Yet these models typically involve homogeneous agents with simple update rules and no optimization objective. Real adaptive systems --- biological, social, economic --- feature heterogeneous interactions and pursue explicit goals such as survival or profit maximization, making optimization under such constraints highly nontrivial. Two questions remain open: can genuinely complex adaptive systems self-organize to criticality without external tuning, and if so, what is the nature of that critical state?

We address both by constructing a minimal model, {\tt EvoSK}, that combines three essential ingredients: (a)~frustrated, heterogeneous interactions via quenched random couplings on a Sherrington--Kirkpatrick (SK) lattice \cite{nishimori2001statistical,garnier2024unlearnable}; (b)~agents that optimize local payoffs through memory-dependent reinforcement learning \cite{garnier2024unlearnable,kidd2012goldilocks,wilson2019eighty}; and (c)~Darwinian selection through extremal replacement of the least-fit agent \cite{bak1993punctuated}. The model extends Bak--Sneppen dynamics \cite{bak1993punctuated} to a setting with adaptive learning on a rugged landscape --- the essential features of living complex systems.

Our analysis yields three results. First, {\tt EvoSK} autonomously reaches a steady state whose collective reward surpasses every non-evolutionary baseline with the same learning rule. Second, avalanche sizes follow a power law with $\tau\approx-1.5$, the mean-field exponent of evolutionary SOC \cite{bak1993punctuated} also observed in neural systems \cite{beggs2003neuronal}. Third, this optimal state sits at the transition between ergodic and non-ergodic phases --- the edge of ergodicity breaking --- identified by a total-variation diagnostic applied to the evolved temperature distribution. We show that these three signatures share a single origin: the branching ratio of the replacement cascade equals the spectral radius of the learning dynamics' Jacobian, so marginal branching and the ergodicity edge are one condition, from which $\tau = -3/2$ follows without adjustable parameters. The critical point is the unique attractor because the selection--mutation balance requires that frozen agents die at a polynomially small rate, which only the critical power-law cascade-size distribution can supply. Each agent's reward is its local stability, so extremal selection repairs the most unstable site; the critical cascades provide the multi-scale rearrangements that effective search requires, explaining why criticality and optimality coincide.

\section{\label{sec:prelim}Model}

\subsection{\label{sec:SKGame} SK Games: Reinforcement Learning on a Spin-Glass Landscape}

The {\tt SK} game \cite{garnier2024unlearnable} is a stylized model of learning in a complex environment. $N$ agents occupy the sites of a fully connected SK lattice in one-to-one correspondence: agent $i$ chooses a binary action $\sigma_i(t) \in \{-1, +1\}$ at each discrete time step $t$, playing the role of the spin at site $i$. All simulations use $N=256$. Interactions are encoded in a quenched random coupling matrix $[J] \in \mathbb{R}^{N\times N}$,
\[
J_{ij} \sim \mathcal{N}(0, 1/N), \quad J_{ii} = 0,
\]
with $J_{ij} = J_{ji}$ (symmetric), which admits an underlying energy function favorable for collective learning \cite{garnier2024unlearnable}.

Each agent maintains action values $q^\sigma_i(t)$ updated by a reinforcement-learning rule,
\begin{equation}
    q^{\sigma}_i(t+1) = (1-\alpha)q^{\sigma}_i(t)+\alpha R^{\sigma}_i(t),
    \label{eq: SK-q-update}
\end{equation}
where $\alpha \in (0, 1)$ is a memory parameter controlling how fast past experience is discounted and $R^{\sigma}_i(t)$ is the instantaneous reward,
\begin{equation}
R^{\sigma}_i(t):=\sigma\sum_{j=1}^{N} J_{ij} \sigma_j(t).
\label{eq: SK-rwd}
\end{equation}
Actions are drawn from a softmax policy,
\begin{equation}
    \mathbb{P}(\sigma_i(t) = \sigma) = \frac{e^{\beta q^{\sigma}_i(t)}}{\sum_{\sigma' = \pm 1} e^{\beta q^{\sigma'}_i(t)}}, \label{eq: SK-act-prob}
\end{equation}
where $\beta=1/T$ is an inverse temperature governing exploration noise.

Equations~(\ref{eq: SK-q-update})--(\ref{eq: SK-act-prob}) define the baseline {\tt SK} dynamics. The parameter $\alpha$ sets effective memory (small $\alpha$: long memory; large $\alpha$: rapid forgetting) and $T$ the exploration level, both fixed externally. All $q_i^\sigma$ start at zero (purely random actions); updates are simultaneous. The dynamics is highly sensitive to both parameters: low $T$ freezes agents into ordered states, high $T$ produces noisy, inefficient exploration.

These dynamics do not reach the SK ground state, and standard $q$-learning convergence guarantees fail: each agent faces a non-stationary environment created by the others' simultaneous learning, $\alpha$ is constant rather than annealed, and the parallel softmax update violates detailed balance. The system instead freezes into one of the exponentially many ``satisficing'' fixed points --- stable under each agent's local rule yet collectively suboptimal \cite{garnier2024unlearnable}. Appendix~\ref{sec:sk-details} gives the full argument. The {\tt SK} game serves as the non-evolutionary reference against which {\tt EvoSK} is benchmarked.

\subsection{\label{sec:evoSK} Evolutionary SK Game with Endogenous Noise Adaptation}

The baseline {\tt SK} game fixes all learning parameters externally. To study how learning interacts with evolutionary selection, we promote the exploration temperature to an evolvable trait. Each agent $i$ receives an individual $T_i$ drawn uniformly from $[0, 2]$, spanning both ergodic and non-ergodic phases of the learning dynamics.

Following the Bak--Sneppen mechanism \cite{bak1993punctuated}, at each time step $t$ agents first update their action values and choose actions via Eqs.~(\ref{eq: SK-q-update})--(\ref{eq: SK-act-prob}). The instantaneous fitness of agent $i$ is
\begin{equation}
f_i(t) := q_i^{\sigma_i(t)}(t),
\end{equation}
the value associated with the action taken. Because both action-value components start and reset at zero and receive opposite-sign rewards, $q_i^{-}(t) = -q_i^{+}(t)$ exactly. The learning state therefore reduces to a single incentive $Q_i := q_i^{+}$, with expected action $m_i = \tanh(Q_i/T_i)$, update $Q_i(t{+}1) = (1{-}\alpha)Q_i(t) + \alpha \sum_j J_{ij}\sigma_j(t)$, and fitness $f_i = \sigma_i Q_i$; we use this reduction throughout. The least-fit agent is then removed and replaced by a blank-slate newcomer,
\begin{equation}
q_{i^*}^{\sigma}(t) = 0 \quad \forall \sigma \in \{\pm 1\},
\end{equation}
whose temperature $T_{i^*}$ is redrawn from $\mathcal{U}[0, 2]$.
This is the Bak--Sneppen convention \cite{bak1993punctuated}: the heritable trait $T_i$ is resampled from a flat kernel, while lifetime knowledge is not transmitted. The flat kernel ensures that any concentration of the surviving $\{T_i\}$ (Sec.~\ref{sec:results-crit-temp}) is produced by selection alone. Appendix~\ref{sec:inherit} confirms that the emergent optimality and edge-seeking are robust to inheritance and parent-centered mutation.
After each replacement, two baseline {\tt SK} updates propagate the newcomer's influence before the next {\tt EvoSK} step (enough for the perturbation to reach every neighbor; results are insensitive to this choice).

\section{\label{sec:results} Emergent Optimality, Criticality, and Avalanches}

\subsection{Evolution-Induced Optimal Learning Performance}

\begin{figure}
\includegraphics[width = \linewidth, trim={.2cm, 0, 0, 0}, clip]{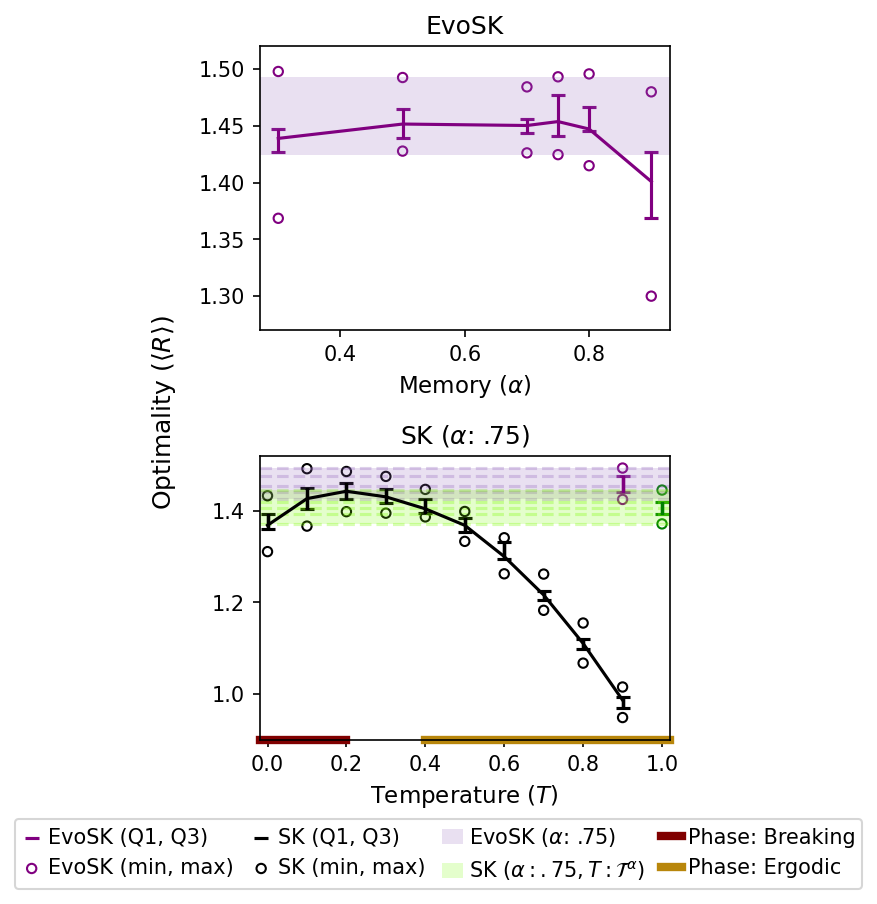}
\caption{\label{fig:opt} \textbf{Optimality.} (a) Steady state rewards $\langle R \rangle$ of {\tt EvoSK} across memory strength $\alpha$ (purple lines), showing a wide span of maxima at $\alpha \approx 0.3-0.8$. This $\alpha$ range mostly coincides with the edge of ergodicity breaking (see Fig.~\ref{fig:wide}). Solid line and error bars indicate the median and interquartile range (Q1--Q3) of $\langle R \rangle$ across the $25$ independent runs at each $\alpha \in \{0.3, 0.5, 0.7, 0.75, 0.8, 0.9\}$ (left- to right-most); open circles mark the minimum and maximum over the same $25$ runs. More extreme memory values ($\alpha = 0.15, 0.95$) are reported in Fig.~\ref{fig:opt-ext}. (b) Black line: $\langle R \rangle$ from {\tt SK} across temperature $T$ at $\alpha = 0.75$.
 Purple shade: $\langle R \rangle$ from {\tt EvoSK} at $\alpha=0.75$. Green shade: {\tt SK} with agents' temperatures assigned to the steady-state temperatures $\mathcal{T}^\alpha: \{T_i\}, t=t_{stat}$ from {\tt EvoSK} runs at $\alpha = 0.75$.
 {\tt EvoSK} outperforms {\tt SK} both at any constant temperature and at the same (frozen) heterogeneous temperatures, indicating that the performance gain stems from the evolutionary dynamics itself.}
\end{figure}

Does the self-organized state confer a functional advantage? Figure~\ref{fig:opt}(a) shows the long-run collective reward $\langle R \rangle$ attained by {\tt EvoSK} across the memory parameter $\alpha$. A broad plateau of near-optimal values spans $\alpha = 0.3$--$0.8$. Beyond this range, performance deteriorates: long-memory agents (small $\alpha$) become trapped in local optima, while short-memory agents (large $\alpha$) over-explore. This is the same range in which the steady state sits at the edge of ergodicity breaking (Sec.~\ref{sec:results-crit-temp}) and produces scale-free avalanches (Sec.~\ref{sec:results-crit-ava}), linking criticality to improved collective outcomes.

At $\alpha=0.75$, {\tt EvoSK} outperforms every constant-temperature baseline --- the {\tt SK} game's ``satisficing solutions'' \cite{garnier2024unlearnable} [Fig.~\ref{fig:opt}(b)]. To test whether the gain comes from the emergent temperature distribution or the evolutionary process itself, we run {\tt SK} with temperatures frozen at the evolved $\{T_i\}$ (green band). This benchmark falls below the constant-$T$ optimum and well below {\tt EvoSK}: the gain is attributable to the dynamics, not to the traits alone. Summing Eq.~\eqref{eq: SK-rwd} over agents gives $\langle R \rangle = -2E/N$, with $E$ the SK energy, so ``collective optimality'' is the canonical hard problem of finding low-energy states of a rugged landscape. The connection to criticality is direct: each agent's reward is its local stability, $R_i = \sigma_i h_i \equiv \lambda_i$ (Appendix~\ref{sec:theory-search}), so the least-fit agent is the least locally stable, and extremal selection is stability repair --- the principle behind extremal optimization~\cite{boettcher2002optimization}. The cascades of Sec.~\ref{sec:results-crit-ava} are therefore cascades of stability repair that span every scale at the critical point, providing the multi-scale rearrangements that effective landscape search requires: below the edge, a greedy quench traps the system in a single basin; above it, agents are chaotic and unable to settle to a good optimum.

\subsection{\label{sec:results-crit-temp} Evolution to the Edge of Ergodicity}

For each $\alpha$, {\tt EvoSK} produces a stationary temperature set $\{T_i\}$ that sets the effective noise of the {\tt SK} dynamics. To locate this state in phase space we evaluate an ergodicity coefficient $\epsilon$ adapted from the Dobrushin coefficient~\cite{seneta1979coefficients, wolfer2024empirical}. $\epsilon$ measures whether independent realizations agree on their macroscopic steady state: we run the non-evolutionary {\tt SK} dynamics under a given $\{T_i\}$ several times, extract each run's steady-state distribution of the collective reward $\bar R(t)=\frac{1}{N}\sum_i R_i(t)$, and take $\epsilon$ as the median total-variation distance between them, Eq.~\eqref{eq: ergo-coeff}. Ergodic dynamics give $\epsilon \approx 0$; broken ergodicity gives $\epsilon \approx 1$. Rescaling all temperatures as $T_i \to T_i 2^\lambda$ then traces a phase profile $\epsilon(\lambda)$ that distinguishes ergodic, non-ergodic, and edge ($0<\epsilon<1$) regimes (Appendix~\ref{sec:perturbEvoSK}).

\begin{figure}
\includegraphics[width = \linewidth]{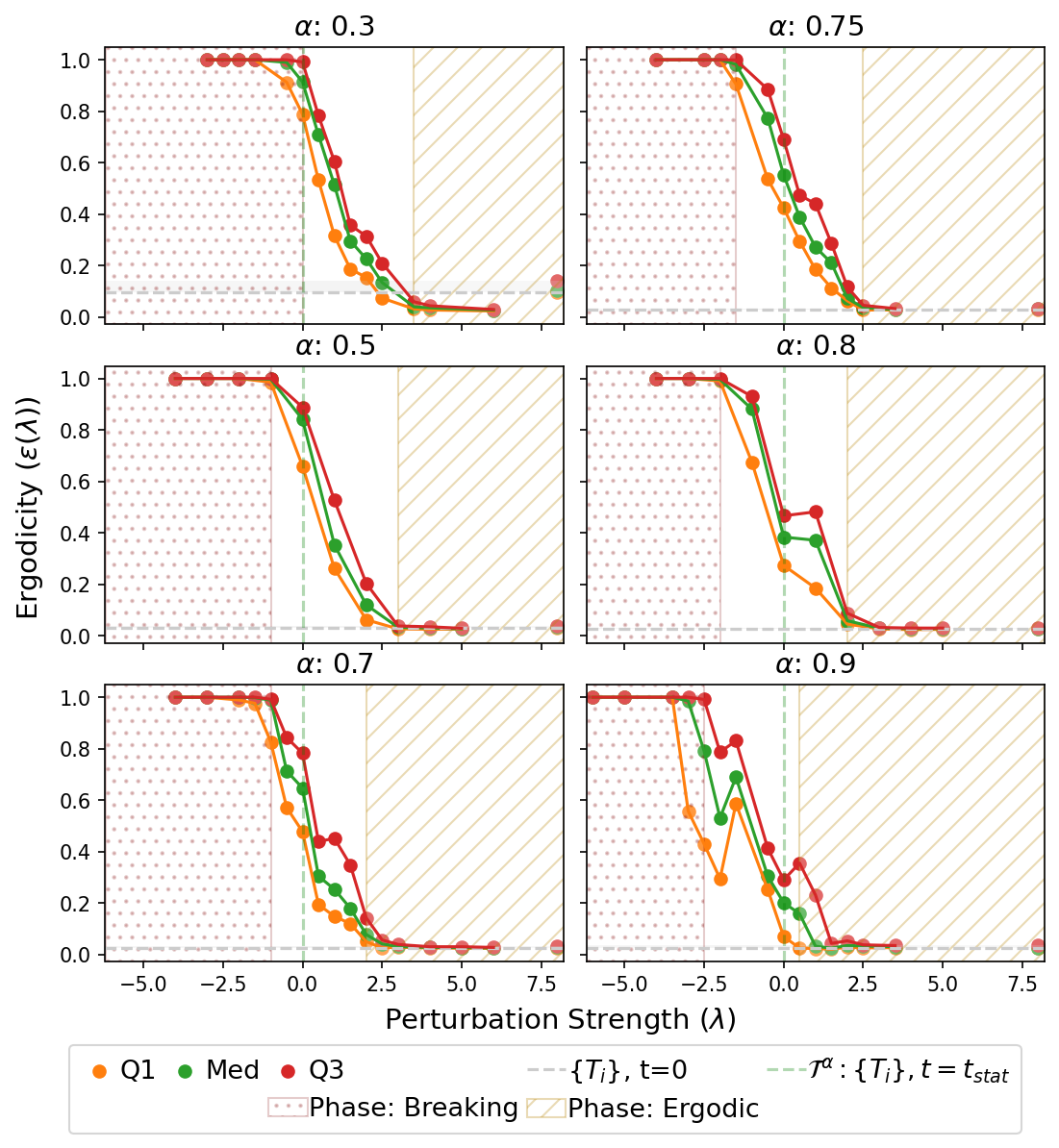}
\caption{\label{fig:wide} \textbf{Ergodicity Phase Diagram.} Panels (a)--(f) correspond to $\alpha = 0.3, 0.5, 0.7, 0.75, 0.8, 0.9$, respectively; each records the ergodicity coefficient $\epsilon$ as a function of the rescaling parameter $\lambda$ (perturbation). At $\lambda=0$, i.e.\ the unperturbed case (green-dotted lines), $\epsilon$ is evaluated on the steady-state agents' temperatures $\{T_i\}$ from $25$ independent {\tt EvoSK} runs at each $\alpha$. Similarly, at other $\lambda$, we have $\epsilon$ associated with each rescaled temperature $\{T_i2^\lambda\}$. At $\alpha=0.3$, $\epsilon(0)$ lies near the breaking phase (red-dotted); at $\alpha=0.9$, $\epsilon(0)$ lies near the ergodic phase (yellow-shaded); at $\alpha=0.5-0.8$, $\epsilon(0)$ lies well within the edge (unshaded). Initial temperatures $\{T_i\}$ at $t=0$ (gray-dotted lines) occupy the ergodic regime across all $\alpha$ values.}
\end{figure}

Figure~\ref{fig:wide} shows $\epsilon(\lambda)$ across memory levels. Over a broad range of $\alpha$ the steady state lies at the edge: at long memory ($\alpha=0.3$) the unperturbed system sits near the ergodicity-breaking phase; at intermediate memory ($\alpha=0.5$--$0.8$) the evolved temperatures fall in the narrow region where $\epsilon$ varies most rapidly; at short memory ($\alpha=0.9$) the system sits near the ergodic phase.

Evolution therefore drives exploration to a phase boundary with no explicit temperature tuning. The initial $\{T_i(0)\}$ places every run in the ergodic phase, so the edge-seeking is produced by replacement and mutation, not inherited from initialization. Appendix~\ref{sec:theory} gives a detailed analysis of the mechanism: an exact selection--mutation balance fixes the evolved density as $\rho^*(T) = u(T)/[N h(T)]$, with $u(T)$ being the (uniform) temperature distribution of newborn agents, and $h(T)$ being the death/hazard rate of an agent at temperature $T$. The parameter-free temperature boundary $T_\times$ defined by $h(T_\times) = 1/N$ coincides with the ergodicity edge at every $\alpha$. That edge is the \emph{critical/marginal-branching} point of the learning dynamics: a single replacement triggers a cascade that heals when the system is below critical branching, and spreads without bound when above it. Populations started hot or in mild cold conditions converge onto this critical point; only a sufficiently deep frozen initial condition remains trapped in a metastable basin, unable to escape.
The critical point is the unique stationary attractor because the balance constrains the cascade-size distribution. Cold agents can die only when an extensive cascade --- one reaching $\mathcal{O}(N)$ flips --- reverses their favorable alignment (``stranding''). Subcritical cascades ($\sigma < 1$) have exponentially bounded sizes, so cold agents are effectively never stranded and the balance fails. Supercritical cascades ($\sigma > 1$) can strand cold agents, but the large warm majority shields them from extremal selection, so the balance again fails. Only near criticality ($\sigma = 1-\delta$, with $\delta\to 0 $ at $N\to\infty$) does the cascade-size distribution $P(S)$ permit \emph{polynomial} weight to extensive events, allowing the system to reach the stationary state meeting the balance. (Appendix~\ref{sec:theory-attract}).

\subsection{\label{sec:results-crit-ava} Critical Avalanche Dynamics Under Evolutionary Pressure}

A second hallmark of edge-seeking dynamics is scale-free cascades. Following Bak and Sneppen~\cite{bak1993punctuated}, we define an avalanche as a contiguous cascade in which agents' fitness falls below a threshold $\delta$ on the drop in the population's minimum fitness, with size $S$ the number of below-threshold agent-updates summed over the avalanche's duration; events are pooled over independent runs (Appendix~\ref{sec:ava-scaling}).

\begin{figure}
\includegraphics[width=.8\linewidth, trim={.2cm .2cm 0 0}, clip]{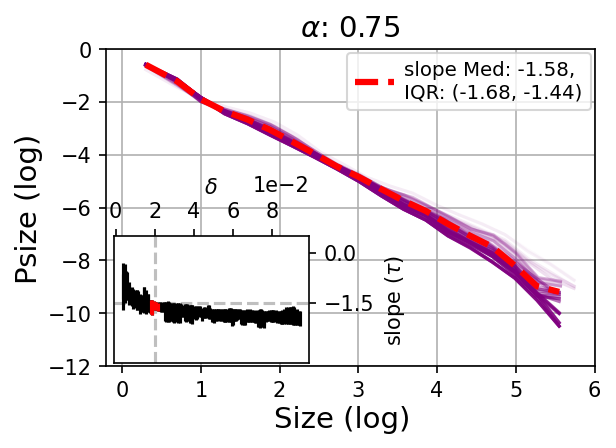}
\caption{\label{fig:epsart-2}
\textbf{Critical Avalanches.} Avalanche size distributions $P(S)$ on log-log axes for $\alpha =0.75$. Distributions are estimated from avalanche events over $25$ runs with $150,000$ time steps each. Each purple line is a $P(S)$ distribution at a given threshold $\delta$; a darker color corresponds to a larger $\delta$. Inset: the interquartile range (IQR) of the local slope $\tau$ across the purple $P(S)$ curves. The red $\delta$-window is where $P(S;\delta)$ stabilizes w.r.t.\ changes in threshold $\delta$.
The corresponding $P(S; \delta)$ is plotted in the main figure as a red dotted line. Exponent $\tau$: median (Q1, Q3) $= -1.58$ ($-1.68$, $-1.44$).}
\end{figure}

Figure~\ref{fig:epsart-2} shows the avalanche size distribution for $\alpha=0.75$: $P(S)$ decays as $S^\tau$ over nearly four decades with $\tau \approx -1.58$, comparable to classical SOC systems. Similar behavior holds for $\alpha=0.5$--$0.8$ (Fig.~\ref{fig:avaCrit-supp}), the same range that reaches the ergodicity edge.

The power-law behavior can be predicted from the branching process underlying the replacement cascade:
once the replacement cascade is identified as a branching process driven to $\sigma = 1$, the total progeny of a critical Galton--Watson process obeys $P(S)\sim S^{-3/2}$, independently of the {\tt SK} substrate or the learning rule (Appendix~\ref{sec:theory-attract}(e)).

Here is how we can extract the critical branching at the ergodicity edge. We can have two copies of the systems, one original and one with an agent flipped action (say be replacing the least fit agent). Under the common-random-number coupling used to measure damage spreading (Appendix~\ref{sec:theory-branch}), an agent acts differently in two copies exactly when the uniform random number separates the two policies --- an event of probability $\tfrac12\chi_i|\delta Q_i|$ --- so the damage process \emph{is} the tangent map of Eqs.~\eqref{eq: SK-q-update}--\eqref{eq: SK-act-prob}. Writing $m_i = \tanh(Q_i/T_i)$ for the expected action and
\begin{equation}
\chi_i \;\equiv\; \frac{\partial m_i}{\partial Q_i} \;=\; \frac{1-m_i^2}{T_i}
\label{eq: chi-main}
\end{equation}
for the agent susceptibility, a perturbation propagates as $\delta Q(t+1) = M\,\delta Q(t)$ with
\begin{equation}
M \;=\; (1-\alpha)\,\mathbb{I} \;+\; \alpha\, J \operatorname{diag}(\chi).
\label{eq: jacobian-main}
\end{equation}
The branching ratio is thus the same quantity as the stability eigenvalue: $\sigma = \rho(M)$, to the accuracy of the annealed approximation ($J$ and $\chi$ averaged together rather than $\chi$ evaluated on the quenched configuration). Because $M$ is the Jacobian at any satisficing fixed point, $\sigma < 1$ makes every such state an attractor and fragments configuration space, whereas $\sigma > 1$ leaves none stable. Critical branching and the ergodicity edge are thus the same condition, and $\tau = -3/2$ is the consequence of such critical branching. Evaluated with no adjustable parameters on the same configurations from which $\sigma$ is measured, $\rho(M)$ crosses unity at $T = 0.28$ against $T = 0.304$ for the damage estimator, and lies inside the $\epsilon(T)$ edge window at every memory level [Appendix~\ref{sec:theory-branch}(e)].

\section{\label{sec:conclude} Conclusions}

We have introduced a minimal model --- {\tt EvoSK} --- in which adaptive learning and evolutionary selection endogenously drive a complex system to the edge of ergodicity breaking. Coupling SK spin-glass dynamics with extremal replacement produces three concurrent signatures of criticality: scale-free avalanches with $\tau \approx -1.5$~\cite{bak1993punctuated}, localization at the ergodic/non-ergodic boundary, and collective rewards surpassing every non-evolutionary benchmark. These three signatures share a single origin: the cascade's branching ratio equals the spectral radius of the Jacobian~\eqref{eq: jacobian-main}, so critical branching and the ergodicity edge are one condition from which $\tau = -3/2$ follows. Criticality is the unique attractor because the selection--mutation balance requires a polynomially small stranding rate for frozen agents, which only the near-critical power-law tail can supply. Moreover, each agent's reward is its local stability, so extremal replacement is stability repair whose scale-free cascades make landscape search most effective --- explaining why criticality, optimality, and self-organization to the edge occur together.

By linking microscopic adaptivity to macroscopic optimality through emergent criticality, our results offer a mechanistic explanation of why diverse biological and socio-economic systems are observed near critical points. Because both criticality and optimality are well defined here, {\tt EvoSK} can serve as a testbed for SOC principles in settings from neural processing to the training dynamics of large-scale learning models.

\begin{acknowledgments}

\end{acknowledgments}

\appendix

\setcounter{equation}{0}
\setcounter{figure}{0}
\renewcommand{\theequation}{S\arabic{equation}}
\renewcommand{\thefigure}{S\arabic{figure}}
\counterwithout*{equation}{section}

\section{\label{sec:method}Method}

At each $\alpha$, randomness enters through (i) the interaction matrix $J$, (ii) the action sampling, and (iii) the initial temperatures $T_i(0)$.
Observables are temperatures $T_i(t)$, rewards $R_i(t)$, and fitness $q_i(t)$ for each agent $i \in [N]$. Steady state is defined by the plateau of the aggregate reward $\bar R(t)$. Because equilibration slows with decreasing $\alpha$ (longer memory), the stationarity onset $t_{\mathrm{stat}}$ is determined separately for each $\alpha$, and all statistics are collected for $t > t_{\mathrm{stat}}$. The total run length of $150{,}000$ steps exceeds the longest equilibration times by an order of magnitude; we verified stability of averages by splitting the post-$t_{\mathrm{stat}}$ window into halves.

\subsection{\label{sec:sk-details}Why the {\tt SK} learning dynamics do not reach the ground state}

Two things prevent the baseline dynamics from relaxing to the ground state of the SK Hamiltonian. First, each agent learns in an environment rendered non-stationary by the simultaneous learning of all others, and the learning rate $\alpha$ is constant rather than annealed, so the stochastic-approximation conditions underlying single-agent $q$-learning convergence theorems \cite{watkins1992q} are violated. Second, and more fundamentally, the coupled dynamics (\ref{eq: SK-q-update})--(\ref{eq: SK-act-prob}) do not obey detailed balance with respect to the global energy $E=-\frac{1}{2}\sum_{ij}J_{ij}\sigma_i\sigma_j$: agents act through a softmax on their accumulated action values --- exponentially weighted averages of \emph{past} rewards --- rather than on the instantaneous local field, and all agents update in parallel. The dynamics is therefore not a Glauber-type relaxation at temperature $T$, and even at low $T$ it need not approach the ground state; instead it freezes into one of the exponentially many ``satisficing'' fixed points, stable under each agent's local update rule yet collectively suboptimal \cite{garnier2024unlearnable}. The ruggedness of the landscape sets the multiplicity of such traps, while the learning rule --- stale estimates, parallel updates, no detailed balance --- is what prevents the system from reaching the deep states that slow sequential annealing could approach \cite{kirkpatrick1983optimization,cugliandolo1994out}.

\subsection{\label{sec:perturbEvoSK} Phase Diagnostics via Temperature Rescaling}

At each $\alpha$, the evolutionary dynamics generate a heterogeneous temperature set
$\mathcal{T}(t;\alpha)=\{T_i(t;\alpha)\}_{i=1}^N$,
which stabilizes after a transient of $t_{\mathrm{stat}}=15{,}000$ steps across all realizations. The emergent temperature ensemble at memory $\alpha$ is
\[
\mathcal{T}^\alpha := \{\mathcal{T}(t_{\mathrm{stat}};\alpha,\omega)\}_{\omega},
\]
where $\omega$ indexes independent realizations.

To probe the dynamical phase we apply a uniform multiplicative perturbation $T_i \mapsto T_i 2^{\lambda}$, $\lambda \in [-6,6]$, preserving the relative structure of $\{T_i\}$ while varying the effective exploration scale, and run the non-evolutionary {\tt SK} dynamics under each rescaled set until stationarity.

Let $\bar R(t)=\frac{1}{N}\sum_i R_i(t)$ be the instantaneous collective reward. From each baseline run at fixed $\lambda$ we extract the empirical steady-state distribution of $\bar R$. Denoting by $\Omega_\lambda$ the set of such distributions across realizations, we define the ergodicity order parameter as
\begin{equation}
\epsilon(\lambda)
:=
\mathrm{median}_{\mu,\nu \in \Omega_\lambda}
\left\| \mu - \nu \right\|_{\mathrm{TV}},
\label{eq: ergo-coeff}
\end{equation}
where the total variation distance between distributions of two realizations $\mu,\nu$ is
\[
\left\| \mu - \nu \right\|_{\mathrm{TV}}
:= \frac{1}{2}\sum_x |\mu(x)-\nu(x)| \in [0,1].
\]
This measure, adapted from the Dobrushin ergodicity coefficient~\cite{seneta1979coefficients, wolfer2024empirical}, quantifies how quickly a Markov chain forgets its initial state. It is computed for each evolutionary realization $\omega$ and summarized across them; its $\lambda$-dependence yields the phase profiles of Fig.~\ref{fig:wide}.

\begin{figure}
    \includegraphics[width =.9\linewidth, trim={.2cm, 0, 0, 0}, clip]{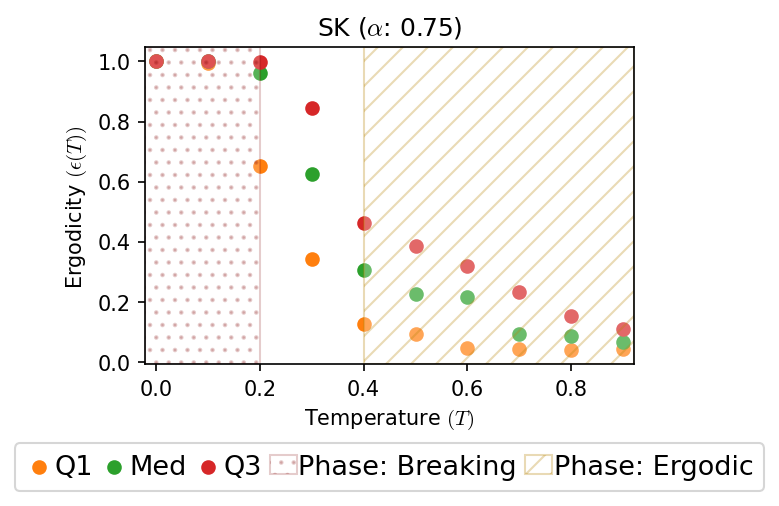}
    \caption{\label{fig:opt-tvdSI}Ergodicity phase diagram of the non-evolutionary {\tt SK} game at $\alpha=0.75$ with uniform, fixed $T$. The same temperature-perturbation analysis used for {\tt EvoSK} applies, with $\epsilon(T)$ replacing $\epsilon(\lambda)$.}
\end{figure}

For baseline {\tt SK} runs at fixed $\alpha, T$, the same coefficient yields $\epsilon(T)$ by replacing $\Omega_\lambda$ with $\Omega_T$; see Fig.~\ref{fig:opt-tvdSI}.

\subsection{\label{sec:ava-scaling} Avalanche Definition and Scaling Estimation}

\begin{figure*}[ht]
\includegraphics[width=\linewidth, trim={.2cm .2cm 0 0}, clip]{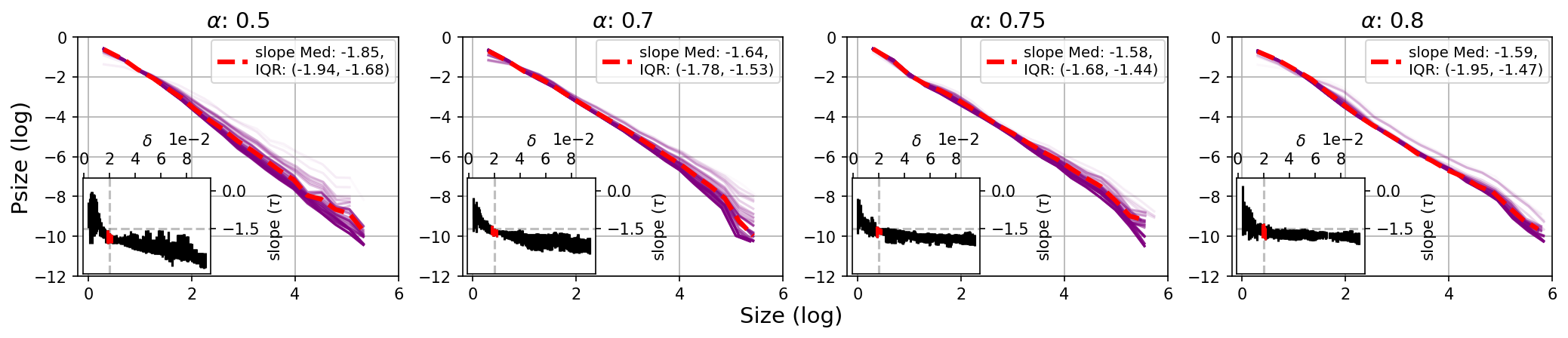}
\caption{\label{fig:avaCrit-supp}
\textbf{Critical Avalanches.} Avalanche size distributions $P(S)$ on log-log axes for $\alpha = 0.5$--$0.8$, coinciding with the clear critical regime (see Fig \ref{fig:wide}). The exponent $\tau$ is flattest and has the smallest slope IQR at $\alpha=0.75$; at $\alpha=0.5$ it steepens to $\tau\approx -1.85$; at $\alpha=0.8$ it remains sufficiently flat but with higher IQR than the rest.
}
\end{figure*}

\begin{figure*}[ht]
\includegraphics[width=0.21\linewidth]{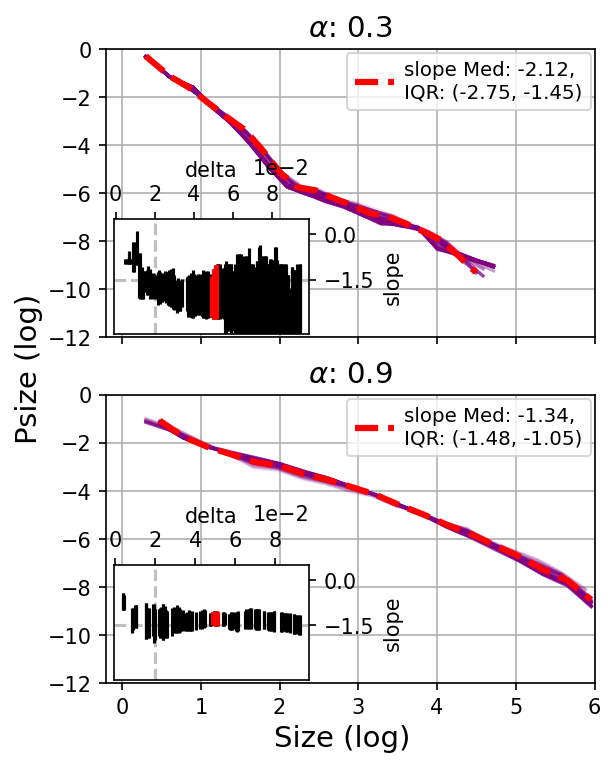}
\includegraphics[width=0.19\linewidth, trim={.5cm, 0, 0, 0}, clip]{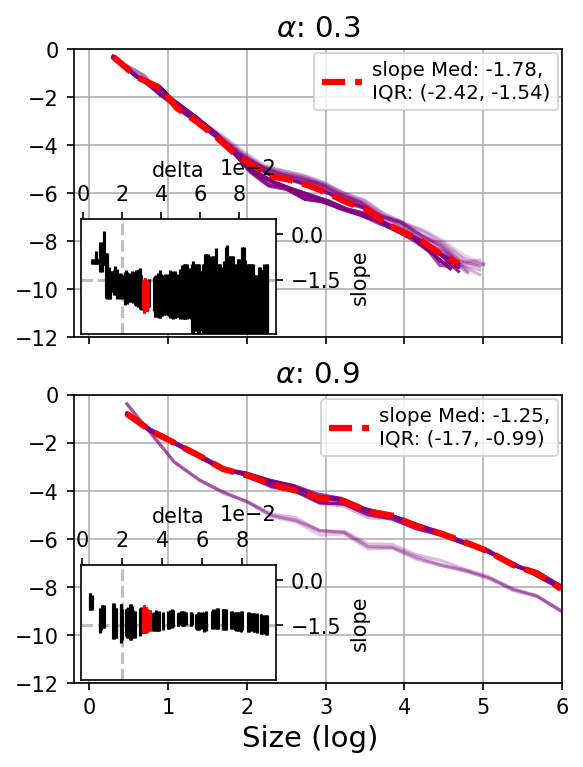}
\includegraphics[width=0.19\linewidth, trim={.5cm, 0, 0, 0}, clip]{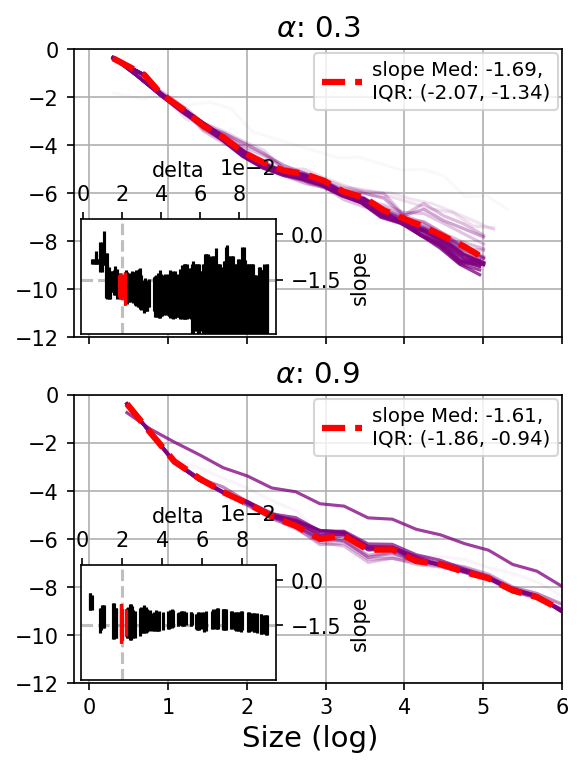}
\includegraphics[width=0.19\linewidth, trim={.5cm, 0, 0, 0}, clip]{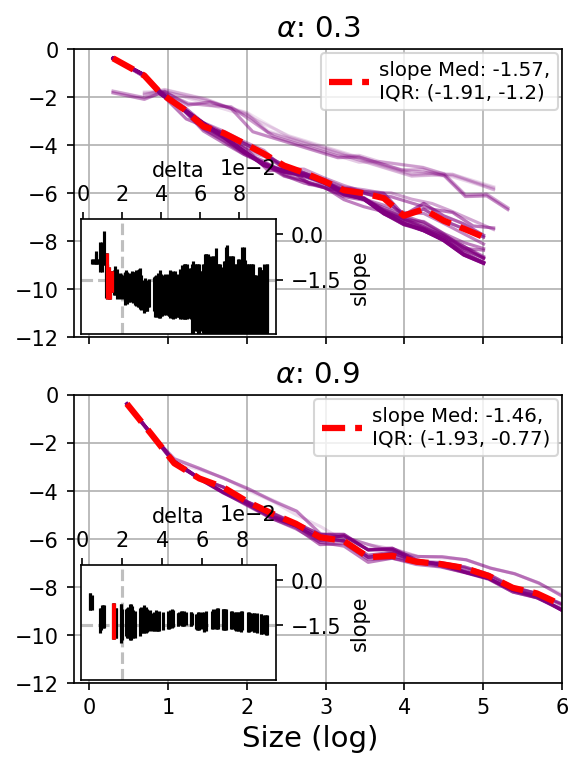}
\includegraphics[width=0.19\linewidth, trim={.5cm, 0, 0, 0}, clip]{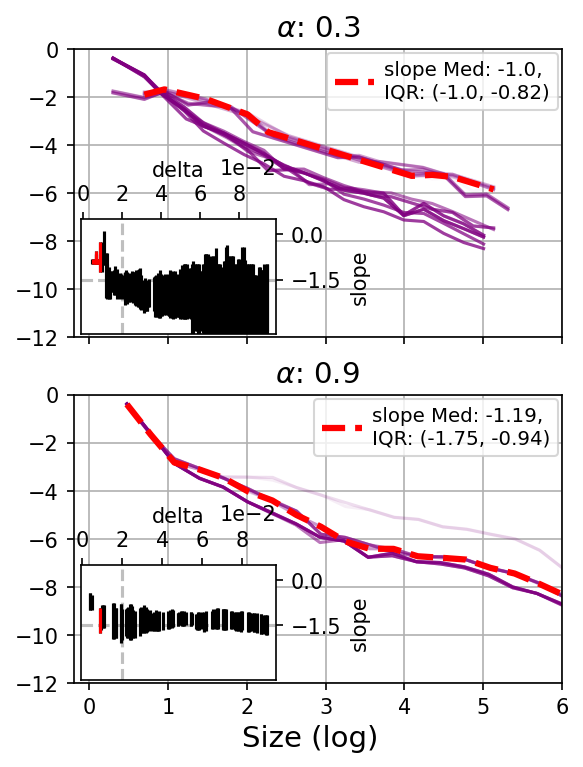}
\caption{\label{fig:wide1}\textbf{Avalanche Distributions Near the Edge.} The first and second rows plot $P(S; \delta)$ across different $\delta$-windows at $\alpha=0.3$ and $\alpha=0.9$, respectively. For both $\alpha=0.3, 0.9$, diagnostics based on local-slope flatness and IQR (see Sec.~\ref{sec:ava-scaling}) find it challenging to identify a consistent power-law regime, in contrast to the intermediate-memory case shown in Fig.~\ref{fig:avaCrit-supp}.}
\end{figure*}

Avalanches are defined by tracking the population's minimal fitness at each step,
$f_{\min}(t) := \min_i f_i(t)$.
Given a threshold $\delta > 0$, an avalanche starts at $t_0$ when
\[
f_{\min}(t_0) < f_{\min}(t_0-1) - \delta,
\]
and ends at the first time $t_1 > t_0$ such that
\[
f_{\min}(t_1) \ge f_{\min}(t_0-1) - \delta.
\]
During an avalanche, agents with fitness below the reference level
$f_{\min}(t_0-1)-\delta$ form a non-empty set
\[
\mathcal{N}(t) := \{ i : f_i(t) < f_{\min}(t_0-1) - \delta \}.
\]
The avalanche size is defined as
\begin{equation}
S := \sum_{t=t_0}^{t_1-1} |\mathcal{N}(t)|.
\end{equation}
The parameter $\delta$ plays the role of the fitness threshold $f_0$ in Bak--Sneppen avalanche definitions~\cite{pruessner2012self}. Because the critical fitness $f_c$ is not directly observable here, $\delta$ provides an operational probe; as in the standard model, scale-free behavior holds only in a restricted threshold window, motivating a systematic scan.

We scan $\delta$ over the dense set of fitness drops
$\{|\Delta f_{\min}(t)| : \Delta f_{\min}(t)<0,\, t>t_{\mathrm{stat}}\}$.
For each $\delta$, avalanche events are pooled across realizations to estimate $P(S;\delta)$. Scaling is assessed via local slopes between consecutive log-binned points of $\log P(S)$ vs.\ $\log S$ (${\sim}\,20$ points per distribution), yielding an ensemble of effective exponents $\tau(\delta)$ summarized by their median and interquartile range (IQR).

The reported exponent is selected where (i) the median local slope is maximally flat, (ii) the IQR is minimal, and (iii) $P(S;\delta)$ is stable under small perturbations of $\delta$. These criteria jointly identify a knee in $\tau(\delta)$ (insets of Fig.~\ref{fig:epsart-2}) that separates a robust scaling regime from threshold choices that distort the statistics.
Supplementary results in Fig. \ref{fig:avaCrit-supp} and Fig. \ref{fig:wide1} show the estimation of exponents for different memory parameter values $\alpha$. The exponent is closest to $-1.5$ and flattest over the widest range of scales for $\alpha = 0.7$--$0.8$, the window in which the collective response of Fig.~\ref{fig:wide} places the system at the edge; away from it the fitted value drifts, reaching $\tau \approx -1.85$ at $\alpha = 0.5$, and at very small or large $\alpha$ (Fig.~\ref{fig:wide1}) $P(S)$ deviates visibly from a power law altogether. We attribute the drift to the finite-size cutoff of $P(S)$ combined with the $\delta$-window selection, both of which steepen the apparent slope when cascades are truncated, rather than to a change of universality, and we therefore quote $\tau$ only over the range in which the fit is stable.

\subsection{\label{sec:branch-est}Branching-Ratio Estimation}

The branching ratio $\sigma$ is estimated from damage-spreading probes as the geometric growth rate of the per-step activity $A(t)$: the number of agents that newly differ between two coupled copies at step $t$. Calibrated on Galton--Watson ensembles of known ratio, the activity fit is unbiased to within $0.035$ over $\sigma = 0.80$--$1.20$ ($0.015$ near criticality).
We fit the ensemble-averaged activity curve rather than averaging per-realization ratios, which are biased low wherever damage dies out.

One thing affects the finite-size scaling of the branching ratio $\sigma$ as shown in Fig.~\ref{fig:branch}(c). The single-flip seed produces a transient damage growth with $N$ per Eq.~\eqref{eq: seed-activity}; including this transient during fitting inflate the large-$N$ estimates to $1.02$--$1.03$. The post-transient fit removes this bias and is stable to the window choice (dropping four or six initial steps agrees to $<0.005$). At the fixed size $N = 256$ used in our main results for all other analysis, the two fitting agree to $<0.01$, and the standard full-window estimator is used.

\subsection{\label{sec:ext-alpha}Extended Memory Range}

\begin{figure}[h]
\includegraphics[width=.85\linewidth]{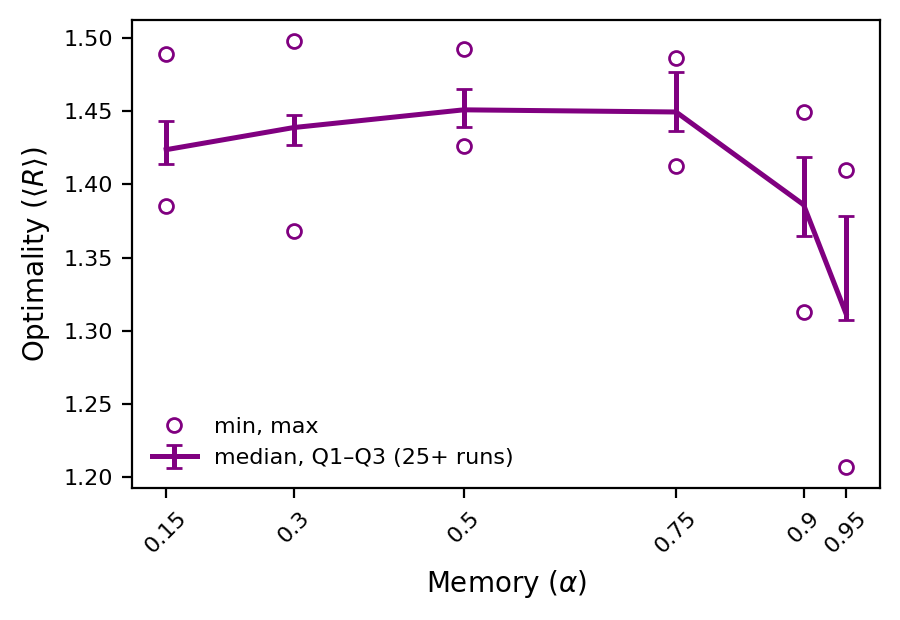}
\caption{\label{fig:opt-ext}\textbf{Steady-state reward at extreme memory values.} Median with interquartile (Q1--Q3) error bars (solid line) and minimum/maximum (open circles) of $\langle R \rangle$ over three independent runs per $\alpha \in \{0.15, 0.3, 0.5, 0.75, 0.9, 0.95\}$, following the conventions of Fig.~\ref{fig:opt}(a), at reduced simulation scale ($5\times 10^4$ steps; steady-state averages over the final $1.5\times10^4$ steps, verified stationary by split-half comparison).}
\end{figure}

Fig.~\ref{fig:opt-ext} extends the reward measurement to more extreme memory values, $\alpha = 0.15$ and $0.95$, using a reduced ensemble (three realizations per $\alpha$, $5\times 10^4$ steps, steady-state averages over the final $1.5\times 10^4$ steps). Intermediate values $\alpha = 0.3, 0.5, 0.75, 0.9$ are re-simulated as consistency anchors and agree with the 25-run medians of Fig.~\ref{fig:opt}(a). The collective reward drops markedly at $\alpha = 0.95$; at $\alpha = 0.15$ the median decreases mildly but with large run-to-run variability, confirming that the near-optimal band does not extend to either extreme of $\alpha$.

\section{\label{sec:inherit}Robustness to Inheritance and Mutation}

The main text initializes every new agent with neutral $q$-values and a temperature from the uniform distribution $\mathcal{U}[0,2]$ --- the Bak--Sneppen convention~\cite{bak1993punctuated}, whose maximum-entropy kernel ensures that any trait concentration reflects selection alone. To establish robustness, we test three biologically motivated alternatives alongside this baseline: \textbf{(V0)} blank-slate (main text); \textbf{(V1)} \emph{elitist inheritance} --- copies both $q$-values and $T$ from the fittest agent; \textbf{(V2)} \emph{fitness-biased inheritance with mutation} --- draws a parent by linear-rank selection, offspring inherits $q$ and $T$ with Gaussian mutation ($\sigma_q = 0.2$, $\sigma_T = 0.1$); \textbf{(V3)} \emph{Darwinian trait inheritance} --- transmits only $T$ from a rank-selected parent ($\sigma_T = 0.1$) while resetting $q$-values. All other parameters match the main text ($\alpha \in \{0.3, 0.5, 0.7, 0.75, 0.8, 0.9\}$; 25 runs of $1.5\times10^5$ steps each).

\begin{figure*}[t]
\includegraphics[width=\linewidth]{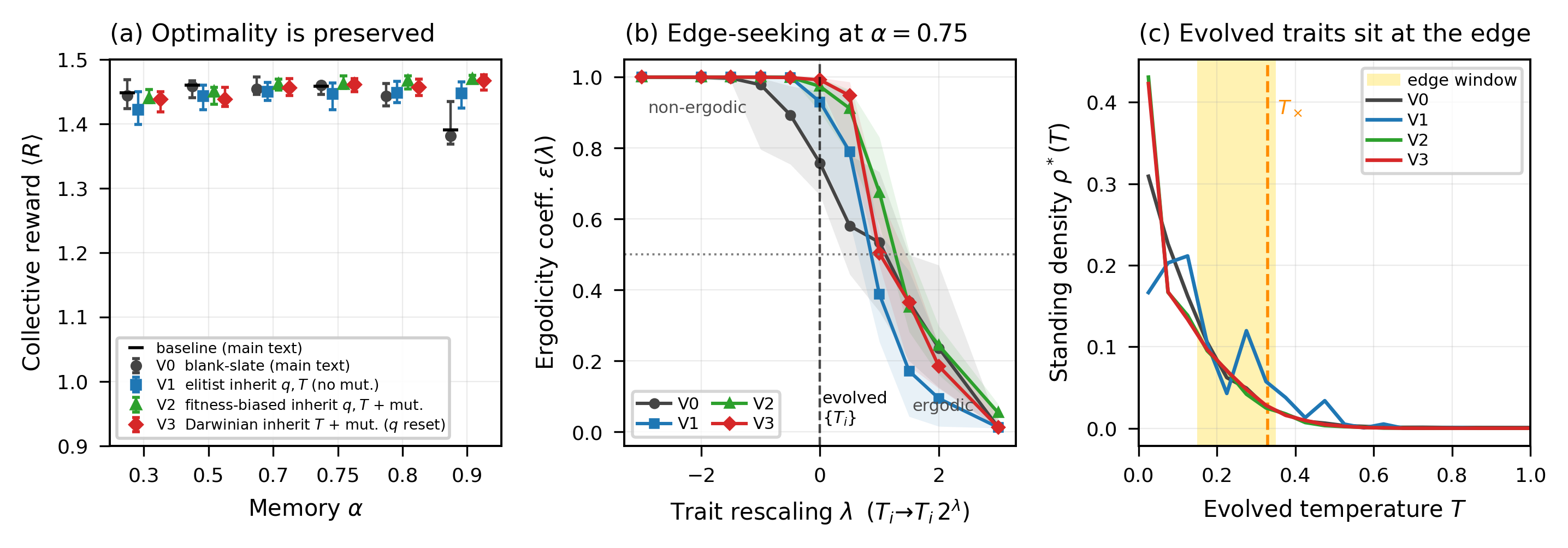}
\caption{\label{fig:inherit}\textbf{Robustness to inheritance and mutation.} Four replacement rules: V0 blank-slate (main text); V1 elitist inheritance of $q,T$ without mutation; V2 fitness-biased inheritance of $q,T$ with mutation; V3 Darwinian inheritance of $T$ with mutation and $q$ reset. \textbf{(a)}~Steady-state collective reward $\langle R \rangle$ vs.\ memory $\alpha$, at the same six $\alpha$ values and the same 25 runs per point (5 disorder samples $\times$ 5 dynamical seeds, $1.5\times10^5$ steps) as Fig.~\ref{fig:opt}(a); median with Q1--Q3 error bars, markers offset horizontally for clarity. Black dashes mark the independently measured baseline medians. Optimality is preserved for all variants across the plateau $\alpha = 0.7$--$0.8$ and improved at $\alpha = 0.9$.
\textbf{(b)}~Ergodicity coefficient $\epsilon(\lambda)$ of the rescaled evolved traits $T_i \to T_i 2^\lambda$ at $\alpha = 0.75$, computed over all 25 evolutionary realizations as in Fig.~\ref{fig:wide} (median; band = Q1--Q3); all variants sit at the same edge (dashed: unperturbed $\lambda = 0$; dotted: $\epsilon = 1/2$). \textbf{(c)}~Evolved standing density $\rho^*(T)$ at $\alpha = 0.75$; shaded: the independently measured edge window; dashed: $T_\times$.}
\end{figure*}

\emph{Optimality is preserved} in every version of the simulations [Fig.~\ref{fig:inherit}(a)]. At $\alpha = 0.75$ the steady-state reward is statistically indistinguishable across all variants (Mann--Whitney test gives $p \geq 0.099$; medians $\langle R \rangle = 1.460, 1.446, 1.462, 1.461$ for V0--V3), and across the plateau $\alpha = 0.7$--$0.8$ no variant differs significantly.
At short memory $\alpha = 0.9$, where the blank-slate model performs worst ($\langle R \rangle = 1.38$), all three inheritance rules \emph{raise} the reward into the optimal band ($\langle R \rangle \approx 1.45$--$1.47$, $p < 10^{-4}$), because a warm start damps the disruptive churn of blank replacement.

\emph{Edge-seeking is preserved.} The temperature-rescaling diagnostic collapses all four rules onto a single ergodicity transition [Fig.~\ref{fig:inherit}(b)]: the unperturbed distribution sits on the non-ergodic shoulder, heating by a factor of two restores ergodicity, and the half-maximum crossing lies within one octave, $\lambda_{1/2} \in [0.86, 1.27]$, for all rules. The evolved density $\rho^*(T)$ concentrates within the same edge window, independently of the inheritance rule [Fig.~\ref{fig:inherit}(c)]. The results suggest that self-organization to the edge is thus a property independent of inheritance, but of extremal selection itself.
This follows from the selection--mutation balance of Appendix~\ref{sec:theory}: the boundary $T_\times$ is set by the extremal hazard, $h(T_\times) = 1/N$ [Eq.~\eqref{eq: Tx-def}], and the mutation kernel controls only the normalization of $\rho^*$, not the edge location.

\section{\label{sec:theory}Theoretical Analysis of the SOC Phenomenon}

Here we provide an in-depth analysis of the theoretical underpinnings of the SOC phenomenon developed in four steps. Section~\ref{sec:theory-exact} derives exact stationarity relations for the evolved temperature distribution and identifies a parameter-free trait-space boundary at the phase edge. Section~\ref{sec:theory-branch} further relates this boundary to the critical/marginal branching of the replacement cascade, whose ratio equals the spectral radius of the learning Jacobian. Damage spreading through incentives and real flips through actions are one operator in two bases, and the resulting branching ratio is the edge of ergodicity, verified empirically across the phase diagram. Section~\ref{sec:theory-attract} explains why evolution is attracted to this marginal point through a stationarity argument, establishes the result in closed form via a solvable two-class caricature, demonstrates evolutionary attraction to this marginality, and derives $\tau = -3/2$ as a theoretical prediction of avalanche size distribution. Section~\ref{sec:theory-search} explains why the self-organized state is also optimal.

Below is the list of important symbols used in this section.

\begin{center}\footnotesize
\begin{tabular}{l p{0.70\linewidth}}
\hline
$\sigma_i$ & action of agent $i$; $\sigma^* = \operatorname{sign}(Q^*)$ a fixed-point configuration \\
$\sigma$ & branching ratio of the damage process; $\sigma_{\mathrm{st}}$ its stationary value in the simplified caricature of Sec.~\ref{sec:theory-attract}(b) \\
$h_i$ & local field $\sum_j J_{ij}\sigma_j$ --- always agent-indexed \\
$h(T)$ & extremal hazard, the per-capita death rate of trait class $T$ --- always a function of $T$, never agent-indexed \\
$\rho^*(T)$ & standing trait density; $\rho(M)$ spectral radius; $\rho^*(Q)$ stationary $Q$-density \\
$\lambda_i$ & local stability $\sigma_i h_i$; $\lambda_0$ the caricature margin of Sec.~\ref{sec:theory-attract}(b); $\lambda$ alone a matrix eigenvalue \\
$\chi_i$ & agent susceptibility $(1-m_i^2)/T_i = \mathrm{sech}^2(Q_i/T_i)/T_i$ \\
$M$ & tangent map in incentive space, Eq.~\eqref{eq: jacobian-main}; $M_\sigma$ its action-space form, Eq.~\eqref{eq: M-sigma}, similar to $M$ \\
$\nu$ & an eigenvalue of $J\operatorname{diag}(\chi)$, so that $M$ has eigenvalues $(1-\alpha)+\alpha\nu$ \\
$u(T)$ & mutation kernel; $u_i$ a leading-eigenvector component \\
$S$ & avalanche size\\
$\mathcal{S}$ & the symmetrised matrix $\chi^{1/2}J\chi^{1/2}$ \\
$R$ & $R_i$ the agent reward and $\langle R\rangle$ the collective reward \\
\hline
\end{tabular}
\end{center}

\subsection{\label{sec:theory-exact}Exact selection--mutation balance and the trait-space boundary at ergodicity edge}

Let $\rho_t(T)$ denote the population density of temperatures at time $t$, normalized as $\int_0^2 \rho_t(T)\,dT = 1$. At each step the global fitness minimum is removed and its temperature resampled from $u(T) = 1/2$ on $[0,2]$. The trait distribution obeys
\begin{equation}
N \rho_{t+1}(T) = N \rho_t(T) - \kappa_t(T) + u(T),
\label{eq: master-rho}
\end{equation}
where $\kappa_t(T)$ is the probability density that the killed agent carries temperature $T$. Stationarity of Eq.~\eqref{eq: master-rho} requires
\begin{equation}
\kappa^*(T) = u(T).
\label{eq: death-balance}
\end{equation}
That is, the death flux must become uniform in $T$, mirroring the flat mutation kernel, even though $\rho^*(T)$ is strongly non-uniform. Writing $\kappa^*(T) = N\rho^*(T)\, h(T)$ with $h(T)$ the per-capita extremal hazard yields
\begin{equation}
\rho^*(T) = \frac{u(T)}{N\, h(T)},
\label{eq: rho-star}
\end{equation}
so temperatures accumulate wherever $h(T)$ is small. Integrating over $T$ gives
\begin{equation}
\int_0^2 \rho^*(T)\, h(T)\, dT = \frac{1}{N},
\label{eq: unit-flux}
\end{equation}
i.e., one death per step. Equations~\eqref{eq: death-balance}--\eqref{eq: unit-flux} are exact for any stationary state and motivate a parameter-free boundary: the crossing temperature $T_\times$ at which
\begin{equation}
h(T_\times) = \frac{1}{N}
\quad\Longleftrightarrow\quad
\rho^*(T_\times) = u(T_\times).
\label{eq: Tx-def}
\end{equation}
Below $T_\times$ the population is \emph{enriched} ($\rho^* > u$); above it the population is \emph{depleted}. The shape of $h(T)$ is measured next and explained in Sec.~\ref{sec:theory-attract}(a), where the hazard floor and enrichment boundary are traced to marginal branching.

\begin{figure*}[t]
\includegraphics[width=\linewidth]{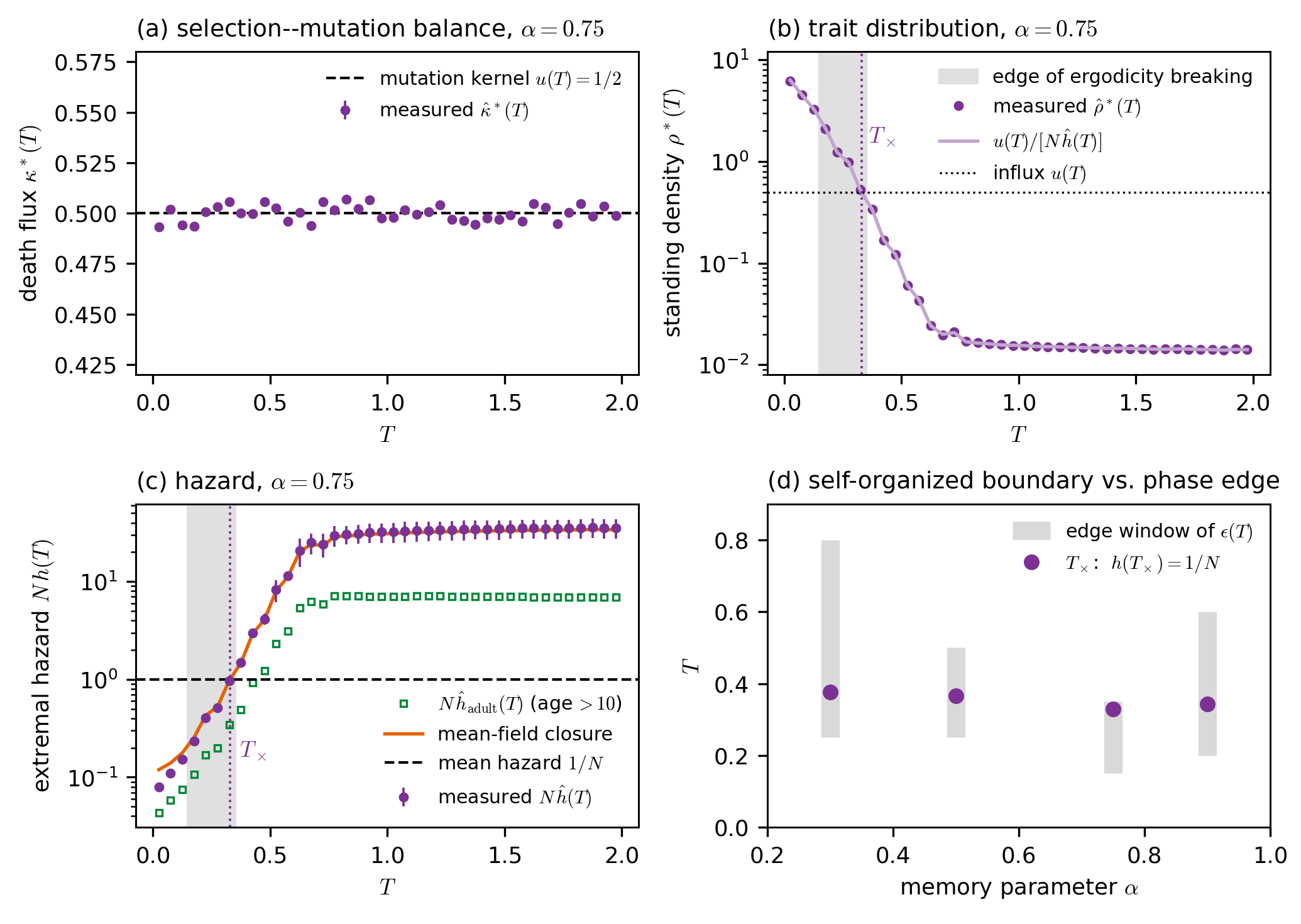}
\caption{\label{fig:theory}\textbf{Numerical verification of the selection--mutation-balance theory.}
(a)~Death-flux density $\hat\kappa^*(T)$ at $\alpha = 0.75$ (25 runs, $3.25\times10^6$ replacement events, error bars: SEM over the 5 disorder realizations), compared with the exact prediction $\kappa^* = u = 1/2$ of Eq.~\eqref{eq: death-balance}.
(b)~Standing temperature density $\hat\rho^*(T)$ (circles) against the balance relation $u(T)/[N\hat h(T)]$ of Eq.~\eqref{eq: rho-star} (line), on a logarithmic scale spanning the $\approx 450$-fold range; dotted line: the flat influx $u(T)$; vertical line: the crossing temperature $T_\times$ of Eq.~\eqref{eq: Tx-def}; gray band: edge window of the ergodicity coefficient $\epsilon(T)$ measured for the constant-$T$ {\tt SK} game at the same $\alpha$ (median $\epsilon$ crossing $0.85 \to 0.15$).
(c)~Extremal hazard $N\hat h(T)$ (circles), its established-agent (age $>10$) component (squares), and the mean-field closure of Eq.~\eqref{eq: closure} (line), which reproduces the measurement with no adjustable parameters; dashed line: the population-mean hazard $1/N$, Eq.~\eqref{eq: unit-flux}.
(d)~The self-organized trait boundary $T_\times$ (circles) falls inside the independently measured edge window of $\epsilon(T)$ (gray bars) at every memory level $\alpha \in \{0.3, 0.5, 0.75, 0.9\}$.}
\end{figure*}

\label{sec:theory-numerics}We verify these relations directly through our simulations. At every post-transient replacement ($t > 2\times 10^4$) we record each agent's temperature (exposure) and the killed agent's temperature, age, and fitness (death flux). Binning $T \in [0,2]$ with $\Delta T = 0.05$ gives estimators $\hat\rho^*(T)$, $\hat\kappa^*(T)$, and $\hat h(T)$. At $\alpha = 0.75$: 25 runs, $1.5\times10^5$ steps, $3.25\times 10^6$ events; at $\alpha \in \{0.3, 0.5, 0.9\}$: 10 runs of $10^5$ steps. Results at $\alpha=0.75$ are summarized in Fig.~\ref{fig:theory}.

\emph{(a) The death flux is uniform.} $\hat\kappa^*(T) = u(T) = 1/2$ to within $1.4\%$ at $\alpha = 0.75$ [Fig.~\ref{fig:theory}(a)] ($\leq 3\%$ at other $\alpha$), while $\rho^*(T)$ varies $\approx 450$-fold. Selection removes agents at a rate mirroring the flat kernel, however non-uniform the population. Small residuals are disorder-specific and do not decay, consistent with glassy relaxation~\cite{cugliandolo1994out}.

\emph{(b) The balance relation holds.} The independent estimates $\hat\rho^*(T)$ and $u(T)/[N \hat h(T)]$ coincide across the full $\approx 450$-fold range [Fig.~\ref{fig:theory}(b)], confirming Eq.~\eqref{eq: rho-star}.

\emph{(c) The hazard has a floor, a knee, and a plateau.} $N\hat h(T)$ rises $\approx 440$-fold across $[0,2]$ [Fig.~\ref{fig:theory}(c)]. As $T \to 0$ it approaches a floor $N h \approx 0.08$: cold agents are not immortal because replacement cascades reshape local fields and strand frozen agents in unfavorable alignments. This floor caps enrichment at $\rho^*(0^+) \approx 6.3$. For $T \gtrsim 0.8$ the hazard saturates at $N h \approx 35$ (mean lifetime $\approx 7$ steps). The knee between these regimes contains the physics: $T_\times = 0.33$ at $\alpha = 0.75$. Decomposing deaths by age shows that $73\%$ of replacements hit ``newborns'' (age $\leq 10$): the extremal replacement propagates through recently replaced agents, as we describe in detail on the avalanche mechanism in Sec.~\ref{sec:results-crit-ava}.

\emph{(d) The hazard obeys a parameter-free extreme-value closure.} If single-site fitness trajectories are statistically independent --- the mean-field factorization justified in the $N \to \infty$ limit of the {\tt SK} game \cite{garnier2024unlearnable} --- then the hazard must obey
\begin{equation}
h(T) = \Big\langle \left[1 - F(f)\right]^{N-1} \Big\rangle_{f \sim p_T},
\label{eq: closure}
\end{equation}
where $p_T(f)$ is the standing fitness density at temperature $T$ and $F(f)$ its population mixture. Evaluated with the \emph{measured} $p_T$ and no adjustable parameters, Eq.~\eqref{eq: closure} reproduces the measured hazard over its full $\approx 440$-fold range at $\alpha = 0.75$ [Fig.~\ref{fig:theory}(c); median ratio $0.98$ over the knee region]. Hence, the shape of the hazard is determined by the shape of the fitness distribution, and the knee is a direct consequence of the extremal-selection mechanism.

\emph{(e) $T_\times$ coincides with the ergodicity edge at every $\alpha$.} We compute $\epsilon(T)$ for the constant-temperature {\tt SK} game [Eq.~\eqref{eq: ergo-coeff}; cf.\ Fig.~\ref{fig:opt-tvdSI}] and define the edge window as the interval where the median $\epsilon(T)$ falls from $0.85$ to $0.15$. $T_\times$ falls \emph{inside} this window at every $\alpha$ tested [Fig.~\ref{fig:theory}(d)]: $T_\times = 0.38, 0.37, 0.33, 0.34$ at $\alpha = 0.3, 0.5, 0.75, 0.9$, against windows $[0.25,0.80]$, $[0.25,0.50]$, $[0.15,0.35]$, $[0.20,0.60]$. The result is robust to threshold choice and stable in $N$: at $\alpha = 0.75$, $T_\times = 0.33(1), 0.33(1), 0.32(1), 0.35(1)$ for $N = 128$--$1024$, with $Nh(T)$ itself $N$-invariant. Equation~\eqref{eq: rho-star} thus implies that, \emph{extremal selection enriches the population on the non-ergodic side up to the phase boundary and depletes it beyond}.

\subsection{\label{sec:theory-branch}What is the edge: the replacement cascade as a branching process}
\label{sec:theory-kernel}

The preceding section establishes \emph{where} the population settles but at the ergordic boundary. It turns out, this boundary is also the marginal branching point, in the process of the system's response to a single replacement.

\emph{(a) The replacement cascade is a branching process.} Replacing the extremal agent likely changes its action, shifting every other local field $h_i = \sum_j J_{ij}\sigma_j$ by $\mathcal{O}(N^{-1/2})$. Agents with $|Q_i| \lesssim T_i$ may flip in response; each flip shifts fields again, and so on. To be more specific, ``in response'' means, under a softmax policy every agent flips spontaneously, so the cascade consists of flips that would not have occurred without the replacement, which is the case of counterfactual realizations measured in (b). The cascade is governed by a branching ratio $\sigma$: the mean flips triggered per flip.
The three regimes ~\cite{zapperi1995self,pruessner2012self} are: $\sigma<1$ --- perturbations heal; $\sigma>1$ --- damage engulfs $\mathcal{O}(N)$ agents; $\sigma=1$ --- scale-free cascades. Structurally this is the mean-field sandpile~\cite{zapperi1995self} and Bak--Sneppen model~\cite{flyvbjerg1993mean}, with agents replacing sites and action flips replacing topplings.

\emph{(b) Measuring $\sigma$: the common-noise coupling.} Two copies run from the same state with identical couplings \emph{and identical action-sampling noise}, differing only in one flipped action at $t=0$; the damage $D(t)$ counts agents whose actions differ. Sharing random numbers removes intrinsic stochasticity, so $D(t)$ tracks exactly the causal descendants of the seed. Sharing the random numbers does more than remove intrinsic noise: it makes the damage process obey the same linearization of the learning dynamics, which lets us compute $\sigma$ rather than only measure it. Under the common-uniform coupling the two copies draw the \emph{same} uniform variate $u_i$, and agent $i$ acts differently in them precisely when $u_i$ falls between the two policies; the probability of that is the length of the interval they span. Writing the policy as $p_i = \mathbb{P}[\sigma_i = +1] = (1+m_i)/2$, so that $\delta p_i = \tfrac{1}{2}\delta m_i = \tfrac{1}{2}\chi_i\,\delta Q_i$ to first order in the incentive difference $\delta Q_i$,
\begin{equation}
\mathbb{P}[\,i \text{ damaged}\,] \;=\; \left| p_i^A - p_i^B \right| \;=\; \tfrac{1}{2}\,\chi_i \left| \delta Q_i \right|,
\label{eq: chi-def}
\end{equation}
with $\chi_i = (1-m_i^2)/T_i$ of Eq.~\eqref{eq: chi-main} and $m_i = \tanh(Q_i/T_i)$ as defined in Sec.~\ref{sec:evoSK}. A damage event flips the action by $\delta\sigma_i = \pm 2$, and averaging over the two signs with the weights just computed gives the mean action difference $\mathbb{E}[\delta\sigma_i] = 2\,\delta p_i = \chi_i\,\delta Q_i$. This closes the loop through the update rule in three steps: the action differences perturb the local fields,
\begin{equation}
\delta h_i \;=\; \sum_j J_{ij}\,\delta\sigma_j \;=\; \sum_j J_{ij}\,\chi_j\,\delta Q_j ,
\label{eq: dh-chain}
\end{equation}
and, substituting Eq.~\eqref{eq: dh-chain}, the fields feed back into the incentives through $Q_i(t+1) = (1-\alpha)Q_i(t) + \alpha\sum_j J_{ij}\sigma_j(t)$, whose linearization is
\begin{equation}
\begin{aligned}
\delta Q_i(t+1) \;&=\; (1-\alpha)\,\delta Q_i(t) + \alpha \sum_j J_{ij}\,\chi_j\,\delta Q_j(t) \\
\;&\equiv\; \big[M\,\delta Q(t)\big]_i ,
\end{aligned}
\label{eq: dQ-map}
\end{equation}
with $M$ exactly the matrix of Eq.~\eqref{eq: jacobian-main}. The damage process is thus described by this linear map, and the mean number of new damage events per existing one is the growth factor of Eq.~\eqref{eq: dQ-map}. The branching ratio $\sigma$ is therefore the spectral radius of the exact Jacobian of the linear map,
\begin{equation}
\sigma \;=\; \rho(M).
\label{eq: sigma-is-rho}
\end{equation}

\emph{(c) The marginal branching ratio is the edge of ergodicity.} Equation~\eqref{eq: sigma-is-rho} is an \emph{annealed} statement: $\rho(M)$ is a one-step growth factor on an instantaneous configuration, whereas $\chi_i(t)$ moves with $Q_i(t)$, so $\rho(M)$ is the annealed counterpart of the maximal Lyapunov exponent of the product $M(t)M(t{-}1)\cdots$ rather than its equal. Nevertheless it characterises the stability of the learning dynamics in the same sense. Evaluated at a satisficing fixed point $Q^* = J\sigma^*$, $\sigma^* = \operatorname{sign}(Q^*)$, $M$ is the Jacobian governing that state's local stability: for $\rho(M) < 1$ every such state is a genuine attractor, configuration space fragments into the exponentially many satisficing fixed points, and independent runs retain their basin --- perturbations heal \emph{and} the ergodicity coefficient satisfies $\epsilon > 0$. For $\rho(M) > 1$ none is an attractor, the trajectory mixes across them, and $\epsilon \to 0$. The identification is structural: for a binary action, $|p_i^A - p_i^B|$ is the total-variation distance $\|\pi_i^A - \pi_i^B\|_{\mathrm{TV}}$ between the two copies' policies at site~$i$, which is Eq.~\eqref{eq: chi-def}. The per-agent damage probability and the ergodicity coefficient $\epsilon$ of Eq.~\eqref{eq: ergo-coeff} are therefore the same kind of object --- a total-variation distance between two copies of the system --- evaluated on one agent's policy over one step in the first case and on the collective reward distribution over an entire run in the second. The ergodicity coefficient and the branching ratio are therefore two readouts of the same eigenvalue crossing unity: the edge of ergodicity breaking \emph{is} the marginal-branching point, $\sigma = 1$.

\emph{(d) Damage and flips: one operator in two bases.}
On the surface, the common-noise coupling of~(b) measures a different object from the cascade of \emph{real} action flips counted by the stationarity argument of Sec.~\ref{sec:theory-attract}(a) and the caricature of Sec.~\ref{sec:theory-attract}(b). However, the two processes share the same operator that determines the branching ratio $\sigma$.

The chain along which a perturbation moves in one step is fixed by the update rule,
\[
\delta\sigma_j \;\xrightarrow{\;J_{ij}\;}\; \delta h_i \;\xrightarrow{\;\alpha\;}\; \delta Q_i \;\xrightarrow{\;\chi_i\;}\; \delta\sigma_i ,
\]
and the two mechanisms --- common-noise coupling and action flips --- simply cut it at different links. Cutting at the incentive gives $M$ of Eq.~\eqref{eq: jacobian-main}, in which $\chi$ is carried by the \emph{source}, since it is what converts that agent's incentive shift into an action shift. Cutting at the action instead --- the natural choice when one is counting flips, as Sec.~\ref{sec:theory-attract}(a) and~(b) do --- gives
\begin{equation}
M_\sigma \;=\; (1-\alpha)\,\mathbb{I} + \alpha \operatorname{diag}(\chi)\,J \;=\; \operatorname{diag}(\chi)\,M\operatorname{diag}(\chi)^{-1},
\label{eq: M-sigma}
\end{equation}
in which $\chi$ is carried by the \emph{target}. The second equality holds because $\delta\sigma = \operatorname{diag}(\chi)\,\delta Q$ to linear order, and $\chi_i > 0$ strictly, so the two matrices are similar and have identical spectra:
\begin{equation}
\rho(M_\sigma) \;=\; \rho(M) \;=\; \sigma .
\label{eq: same-spectrum}
\end{equation}
Therefore, the branching ratio is the same whether one counts damage or flips, and the two processes are equivalent in the sense that they share the same multiplier $\sigma$.

Note that all these assume the cascades are in the dilute regime $D \ll N$ in which $\sigma$ is fitted, so that linearization of the dynamics holds; beyond it the damage process and the real cascade leave the linear regime together, which is why the supercritical branch saturates at $\mathcal{O}(N)$.

In practice we extract $\sigma$ from the geometric growth rate, in the dilute regime $D \ll N$, of the per-step \emph{activity} $A(t)$ --- the number of agents that \emph{newly} begin to differ at step $t$ --- which for a branching process obeys $\langle A(t)\rangle \propto \sigma^t$. Its calibration and mechanisms are documented in Appendix~\ref{sec:branch-est}; each point pools $200$--$400$ probes over $24$ disorder realizations.

\emph{(e) Empirical verification: the edge is the marginal-branching point.} Figure~\ref{fig:branch}(a) shows $\sigma(T)$ for the non-evolutionary game at uniform $T$ and $\alpha=0.75$. It rises from $\sigma = 0.70$ at $T=0.05$ to $1.28$ at $T=1.4$, crossing unity at $T = 0.304$ ($68\%$ CI $[0.297,0.309]$). The two limits are clear [Fig.~\ref{fig:branch}(b)]: at $T=0.10$ damage decays to zero; at $T=1.0$ it grows and saturates at $\mathcal{O}(N)$. The crossing lies \emph{inside} the $\epsilon(T)$ edge window $[0.15,0.35]$ and adjacent to $T_\times = 0.329$.
The co-location is structural [cf.\ (c) above]: the same branching/Lyapunov threshold governs the proliferation of dynamically accessible equilibria in random dynamical systems~\cite{wainrib2013topological,fyodorov2016nonlinear}, and $\sigma = 1$ is the edge-of-chaos condition~\cite{sompolinsky1988chaos}.
We verify Eq.~\eqref{eq: sigma-is-rho} directly, evaluating $\rho(M)$ on the \emph{same} configurations from which $\sigma$ was measured, paired sample by sample. The two curves agree to within $0.036$ across the entire edge region, and $\rho(M)$ crosses unity at $T = 0.28$ against $T = 0.304$ for the damage estimator [Fig.~\ref{fig:branch}(a); same $13$ temperatures, $48$ disorder realizations each].

This identification holds across the phase diagram, not only at $\alpha = 0.75$. Repeating the $\sigma(T)$ sweep at every memory level, the marginal crossing falls inside the independently measured $\epsilon(T)$ edge window in each case: $T(\sigma{=}1) = 0.363, 0.361, 0.304, 0.256$ against edge windows $[0.25,0.80]$, $[0.25,0.50]$, $[0.15,0.35]$, $[0.20,0.60]$ at $\alpha = 0.3, 0.5, 0.75, 0.9$, and adjacent to the balance boundaries $T_\times = 0.376, 0.366, 0.329, 0.344$ of Sec.~\ref{sec:theory-numerics}. The parameter-free criterion $\rho(M) = 1$ falls inside the same four windows ($T = 0.74, 0.43, 0.28, 0.24$).
The edge of ergodicity breaking, the marginal-branching point, and the trait-space boundary $T_\times$ thus coincide at every $\alpha$ tested, so the mechanism below is a property of the model rather than of one parameter value.

\subsection{\label{sec:theory-attract}The marginal-branching point, and why evolution is driven to it}

\emph{(a) Why evolution is attracted to $\sigma=1$: a stationarity argument.} The attractor follows from the exact relations of Sec.~\ref{sec:theory-exact} alone. Equation~\eqref{eq: death-balance} demands that \emph{every arrival dies exactly once}, so the kill process must visit every trait class at rate $u(T)$ --- including the coldest. One caveat: a state where \emph{every} arrival is culled in infancy satisfies $\kappa^*(T) = u(T)$ trivially, since newborn traits are drawn from $u$ by construction. This degenerate solution is the frozen branch of~(c) below. The question is which dynamical regime, if any, sustains turnover among established agents.

Cold agents die by \emph{stranding}: a replacement cascade reshapes local fields and reverses a frozen agent's favorable alignment, making it vulnerable to extremal selection [Sec.~\ref{sec:theory-exact}(c)]. A cascade of $S$ flips displaces the local field of a non-participating agent~$i$ by $\delta Q_i = \sum_{j \in \text{cascade}} J_{ij}\,\delta\sigma_j$. Each flip contributes $\delta\sigma_j = \pm 2$ and each coupling has variance $1/N$, so $\operatorname{Var}(\delta Q_i) = 4S/N$. A displacement large enough to reverse the agent's alignment therefore requires $S$ of order~$N$: only an \emph{extensive} cascade can strand a cold agent. But stranding is necessary, not sufficient: the stranded agent must also own the global fitness minimum to be removed by the extremal rule. Cold-agent death is therefore a two-step process --- an extensive cascade to strand, then extremal selection to kill --- and the three regimes differ in which step, if either, succeeds.

\emph{Subcritical} ($\sigma < 1$): cascade sizes are exponentially distributed, $P(S \geq s) \sim e^{-s/s_c}$ with $s_c$ finite. The probability of an extensive cascade is exponentially small in $N$, so the first step already fails: cold agents are effectively never stranded, and the question of selection never arises. The death flux fails Eq.~\eqref{eq: death-balance} classwise, and the state remains trapped in the frozen basin of~(c) below.

\emph{Supercritical} ($\sigma > 1$): cascades bifurcate --- each either dies quickly or survives to engulf $\mathcal{O}(N)$ agents --- with no intermediate power-law regime. The surviving cascades \emph{can} strand cold agents, so the first step succeeds; but the second step fails. The field-perturbation variance per step is $\operatorname{Var}(\delta Q) = 4\langle S\rangle/N$; supercritically $\langle S\rangle = \Theta(N)$, so $\operatorname{Var}(\delta Q) = \Theta(1)$ and each of the $\Theta(N)$ warm agents makes $\mathcal{O}(1)$ fitness excursions through zero. At any step a fraction $p = \mathcal{O}(1)$ of them sit at negative fitness. A stranded cold agent sits at $f = -\mathcal{O}(\lambda_0)$ --- the overshoot past the stability gap --- a fixed $\mathcal{O}(1)$ depth. The extreme value over $\Theta(N)$ warm agents, each independently sampling $\mathcal{O}(1)$ excursions, is deeper than this fixed level with probability $1 - (1-p)^{\Theta(N)}$, exponentially close to one. Therefore $\kappa(\text{cold}) \to 0 \neq u$: stranding occurs but selection does not, and the state is again not stationary. The drift out of this regime has definite sign: warm agents are selected and die at the normal rate, while cold agents are shielded and survive. Since only warm agents can flip and propagate cascades, their depletion lowers $\sigma$ --- a supercritical population cools until it is not.

\emph{Marginal} ($\sigma = 1$): cascades are critical branching processes with the universal size distribution $P(S) \sim S^{-3/2}$. Unlike the subcritical case, this power-law tail assigns polynomial weight to extensive cascades: $P(S \geq cN) \sim (cN)^{-1/2}$. The stranding probability for a cold agent is therefore of order $N^{-1/2}$ per cascade --- polynomially small rather than exponentially small. The selection step now operates as well: at criticality $\langle S\rangle = \Theta(\sqrt{N})$, so the field-perturbation variance per step drops to $\Theta(N^{-1/2})$ --- warm agents' fitness fluctuations are now parametrically smaller than the stranded cold agent's depth $\mathcal{O}(\lambda_0)$, so a stranded cold agent is the clear fitness minimum [quantified in~(b)]; the cold-agent hazard is controlled by the stranding rate, $h_C \propto \langle\varphi_N\rangle$, which matches the measured floor $Nh \approx 0.08$ [Sec.~\ref{sec:theory-exact}(c)]. The precise balance, which selects the finite-size correction $1 - \sigma \sim N^{-1/2}$, is derived in part~(b). Marginal transport is the unique regime in which the balance conditions are satisfiable with non-vanishing established-agent turnover; by Eq.~\eqref{eq: sigma-is-rho} it is the ergodicity edge. In the Bak--Sneppen model~\cite{bak1993punctuated,paczuski1996avalanche}, the system self-organizes to an emergent fitness threshold: species below it are replaced, and the steady state pins the population minimum at that threshold. Here the self-organized threshold is not a fitness value but a phase boundary of the substrate's dynamics --- on one side cascades cannot strand cold agents, on the other warm agents' $\mathcal{O}(1)$ fitness excursions shield them from selection. The population is driven to the unique point between these two regimes: the branching transition $\sigma = 1$, the edge of ergodicity breaking. The self-tuning mechanism parallels adaptive criticality via depleting synapses~\cite{levina2007phase} and the marginal-stability principle in glasses~\cite{muller2015marginal}, with extremal selection playing the role of the feedback that pins the system at the phase boundary.

\label{sec:theory-caricature}
\emph{(b) The three regimes in closed form.} To make the preceding argument quantitative we strip the model to three ingredients --- extremal removal, a mutation kernel with full support, and a hard class that only collective events can kill --- discarding the {\tt SK} substrate, the learning rule, and the continuous trait.

Cold agents freeze into deep alignment with their local field: from Eq.~\eqref{eq: chi-main}, $\chi = (1-m^2)/T$ vanishes exponentially once the stability margin $\lambda$ exceeds~$T$, because $m = \tanh(\lambda/T)$ saturates to $\pm 1$. Each cold agent therefore sits at a stability margin $\lambda \geq \lambda_0 > 0$, with $\lambda_0$ the minimum gap between a frozen agent's local field and its flipping threshold; they neither propagate cascades nor flip spontaneously. Warm agents, by contrast, retain susceptibility $\chi$ of order $1/T$: they respond to perturbations, propagate cascades, and make $\mathcal{O}(1)$ fitness excursions through zero. The only death channel for a cold agent is therefore stranding: a cascade of warm-agent flips must displace the cold agent's field by more than $\lambda_0$.

Take $N$ agents, each \emph{cold} or \emph{warm}, with warm fraction $w$. One agent --- the least fit --- is removed per step and replaced by a newborn drawn warm with probability $\beta \in (0,1)$, independently of the state, so the kernel has full support on both types. Warm agents are \emph{soft}: they carry the cascades, so the branching ratio is proportional to their share,
\begin{equation}
\sigma \;=\; a\,w ,
\label{eq: caric-sigma}
\end{equation}
with $a$ a constant set by the substrate's susceptibility. A fully warm population ($w = 1$) is supercritical, so $a > 1$; this is what makes marginality attainable at all: it puts the critical warm fraction at $w_* = 1/a < 1$, so a marginal population can still hold an $\mathcal{O}(1)$ cold majority --- the enriched state Eq.~\eqref{eq: rho-star} describes. Cold agents can die only by being \emph{stranded}, as established above. Acquiring the margin $\lambda_0$ at birth is what protects cold newborns from immediate culling, and so excludes the degenerate branch set aside in~(a).

Every arrival dies exactly once, so cold agents must die at precisely the rate the kernel injects them,
\begin{equation}
N(1-w)\,h_C \;=\; 1-\beta ,
\label{eq: cold-balance}
\end{equation}
a cold mortality of order $1/N$ --- polynomially small, but not smaller. That scale is the crux of the argument. As in part~(a), a cascade of $S$ flips displaces a cold agent's field by $\Delta h \sim \mathcal{N}(0, 4S/N)$. Stranding needs $|\Delta h| > \lambda_0$, so the Gaussian tail gives the stranding probability
\begin{equation}
\varphi_N(S) \;=\; \mathbb{P}\big[\,|\Delta h| > \lambda_0\,\big]
\;\simeq\; e^{-S_*/S},
\qquad S_* \;\equiv\; \lambda_0^{2}N/8 ,
\label{eq: strand-kernel}
\end{equation}
which is smaller than any power of $N$ unless the cascade reaches the \emph{extensive} scale $S_*$: a cold agent cannot be picked off by a local event, only by a collective one. The cold-agent hazard is set by averaging this stranding kernel over the cascade-size distribution. With the Galton--Watson progeny law $P(S) = C S^{-3/2}e^{-S/s_c}$, and cutoff $s_c = A(1-\sigma)^{-2}$. The amplitude $A$ is not free: the exact Galton--Watson cutoff is $1/s_c = \sigma - 1 - \ln\sigma$, which near $\sigma = 1$ expands to $(1-\sigma)^2/2$ and therefore fixes $A = 2$. The average is then closed-form,
\begin{equation}
\big\langle \varphi_N \big\rangle_S
\;=\; C\sqrt{\frac{\pi}{S_*}}\;e^{-2\sqrt{S_*/s_c}} ,
\label{eq: caric-integral}
\end{equation}
by the elementary integral $\int_0^\infty S^{-3/2}e^{-aS-b/S}\,dS = \sqrt{\pi/b}\,e^{-2\sqrt{ab}}$. Two features carry everything. The prefactor is $\Theta(N^{-1/2})$ for \emph{every} $\sigma$, since $S_* \propto N$. And all remaining $\sigma$- and $N$-dependence enters the exponent through a single scaling variable $z \equiv (1-\sigma)\sqrt{N}$, because $S_*$ carries the whole $N$-dependence and $s_c$ the whole $\sigma$-dependence:
\begin{equation}
2\sqrt{S_*/s_c} \;=\; 2\sqrt{\frac{\lambda_0^2 N}{8}\;\frac{(1-\sigma)^2}{2}}
\;=\; \frac{\lambda_0\,z}{2} .
\label{eq: zvar}
\end{equation}
Stranding is necessary but not sufficient: the stranded agent must also own the global fitness minimum. The cold-agent hazard therefore factorizes as $h_C = \langle\varphi_N\rangle \times P_{\mathrm{sel}}$. At the balance point derived below, $n_{\mathrm{str}} = N(1-w)\langle\varphi_N\rangle = \Theta(1)$: stranding is intermittent. When a cold agent is stranded, its fitness sits at $-\mathcal{O}(\lambda_0)$ (the overshoot past the stability gap), while warm agents' per-step field perturbations have amplitude $\mathcal{O}(N^{-1/4})$ at criticality --- too small to reach this depth at large~$N$ (at the simulation size $N = 256$, $N^{-1/4} = 0.25$ is comparable to the overshoot, so this separation is asymptotic). The stranded cold agent is therefore the clear minimum and is selected within $\mathcal{O}(1)$ steps: $P_{\mathrm{sel}} = \mathcal{O}(1)$. With $P_{\mathrm{sel}}$ absorbed into the $\mathcal{O}(1)$ prefactor, the balance reduces to $h_C \propto \langle\varphi_N\rangle$.

Setting $h_C \propto \langle\varphi_N\rangle$ equal to the $\Theta(1/N)$ demanded by Eq.~\eqref{eq: cold-balance} and dividing out the common $N^{-1/2}$ prefactor, the entire balance condition reduces to one scalar equation,
\begin{equation}
\sqrt{N}\;e^{-\lambda_0 z/2} \;=\; \Theta(1) ,
\label{eq: caric-master}
\end{equation}
and the three transport regimes are the three answers for where $z$ must sit.

Independent warm-agent flips also perturb a cold agent's field, but they do not accumulate into a persistent displacement. Each warm agent has $m_j \approx 0$ and flips stochastically between orientations, contributing an oscillating signal to the cold agent's local field $\bar{h}_i = \sum_j J_{ij} m_j$ that averages to zero. The persistent field --- the quantity the cold agent's stability is measured against --- shifts only when a neighboring agent permanently changes sign. A critical cascade concentrates $\Theta(N)$ correlated, permanent flips into a single event [Eq.~\eqref{eq: strand-kernel}], making cascade-mediated stranding the mechanism that displaces the persistent field past~$\lambda_0$.

\emph{Subcritical:} hold $\sigma \leq 1-\delta$ fixed as $N$ grows. Then $z = \delta\sqrt N$ diverges and the left side of Eq.~\eqref{eq: caric-master} is $\sqrt{N}e^{-\Theta(\sqrt N)} \to 0$ --- the localization of~(a) in closed form, i.e. effectively cold agents do not get stranded.

\emph{Supercritical:} now the stranding probability is $\Theta(1)$ per cascade, so cold-agent strandings are amply supplied. But the obstruction to cold agents' replacements moves to the extremal selection rule: $\Theta(N)$ warm agents make $\mathcal{O}(1)$ negative fitness excursions, while cold agents sit at $\lambda \geq \lambda_0 > 0$. The global minimum is therefore warm with probability exponentially close to one, giving $\kappa_C \to 0$: cold agents stop dying even as they get stranded. The two regimes fail for different reasons --- one by what cascades can reach, the other by what the extremal rule can select.

\emph{Marginal:} what remains is the neighbourhood of $\sigma = 1$, and not $\sigma = 1$ exactly. At $z = 0$, $n_{\mathrm{str}} = \Theta(\sqrt{N})$: a strictly critical population would strand $\sqrt{N}$ cold agents per step, but only one can die per step. Cold agents would permanently occupy the extremal minimum, shutting out warm-agent deaths and violating the warm-agent balance. Since the left side falls monotonically from $\sqrt N$ to zero as $z$ increases, it crosses $\Theta(1)$ exactly once, at $\lambda_0 z/2 = \tfrac12\ln N$, giving the stationary branching ratio:
\begin{equation}
1 - \sigma_{\mathrm{st}} \;\simeq\; \frac{\ln N}{\lambda_0\sqrt{N}} ,
\label{eq: sigma-star}
\end{equation}
approached \emph{from below}: slightly subcritical at any finite size, but at criticality in the thermodynamic limit. So the balance condition is satisfied only in the marginal regime, and the population is driven to the edge of ergodicity breaking.

\begin{figure*}[t]
\includegraphics[width=.9\linewidth]{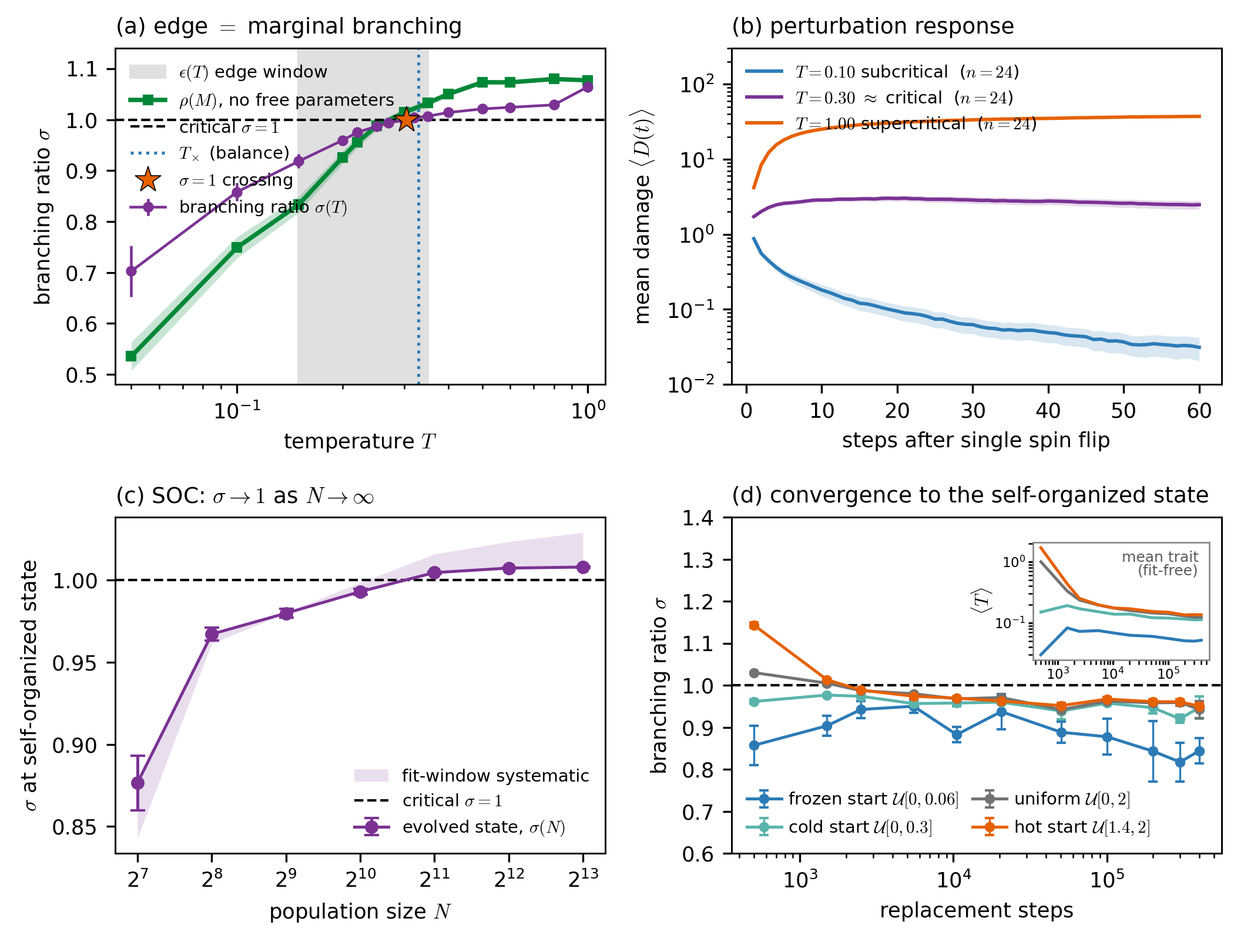}
\caption{\label{fig:branch}\textbf{{\tt EvoSK} self-organizes to the critical-branching point.}
(a)~Branching ratio $\sigma$ of the replacement cascade, measured by damage spreading on the non-evolutionary game at uniform temperature ($\alpha = 0.75$; $200$--$400$ probes per point, pooled over $24$ disorder realizations; bootstrap SE over realizations). $\sigma$ crosses unity at $T = 0.304$ (star; $68\%$ CI $[0.297,0.309]$), inside the independently measured $\epsilon(T)$ edge window (shaded) and adjacent to the balance boundary $T_\times$ (dotted). Green line: the parameter-free prediction $\rho(M)$ of Eq.~\eqref{eq: sigma-is-rho}, evaluated at the same temperatures and on the same configurations as $\sigma$ ($48$ disorder realizations each; band is the SEM), which crosses unity at $T = 0.28$. The range is truncated at $T = 1.0$: above it the activity fit inflates while $\rho(M)$ turns over, the estimator bias discussed in Sec.~\ref{sec:theory-branch}(e). Perturbations heal in the frozen population and spread in the chaotic one.
(b)~Mean damage $\langle D(t)\rangle$ after a single action flip, for a subcritical, a near-critical, and a supercritical temperature; shaded bands are $\pm 1$ SEM over $24$ disorder realizations.
(c)~Branching ratio of the \emph{evolved} trait set as a function of population size, from $N=128$ to $N=8192$ ($24$ disorder realizations per size, $8$ at $N=8192$; $N=64$ excluded per Appendix~\ref{sec:branch-est}). The solid line is the transient-free estimate --- the reliable branching ratio; error bars are the bootstrap SE over realizations: the self-organized state approaches $\sigma = 1$ from below and \emph{saturates} at $\sigma \approx 1.01$ beyond $N \approx 10^3$, the finite-size signature of a self-organized branching process \cite{zapperi1995self}. The shaded band is the fit-window systematic, spanning up to the full-window fit, which the early-time seed transient inflates to $\approx 1.03$ at the largest $N$ [Sec.~\ref{sec:theory-attract}(d)].
(d)~Convergence to the self-organized state. Starting the population cold, at the mutation kernel $\mathcal{U}[0,2]$, or hot ($\sigma = 1.14$), the evolutionary dynamics drives all three to a common band, $\sigma \approx 0.95$, consistent with the $N = 256$ post-transient value $\sigma = 0.97$ of panel~(c), the residual being small against the realization scatter of the convergence protocol, over $4\times10^5$ replacement steps (pooled over $24$ disorder realizations; error bars are the bootstrap SE over realizations, as in panels (a) and (c)). The frozen start ($\mathcal{U}[0,0.06]$) is the exception: it remains in a separate, colder metastable basin ($\sigma \approx 0.85$, $\langle T\rangle \approx 0.05$), the dynamical signature of the broken ergodicity of the frozen phase [text, Sec.~\ref{sec:theory-attract}(c)]. Inset: the mean trait $\langle T\rangle(t)$ (fit-free, hence far less noisy than $\sigma$), showing the same convergence and the same frozen exception.}
\end{figure*}

\emph{(c) The edge is approached, not inherited.} We initialize the population in frozen ($T_i \sim \mathcal{U}[0,0.06]$, $\sigma = 0.86$), slightly cold ($\mathcal{U}[0,0.3]$), slightly warm (same as the mutation kernel $\mathcal{U}[0,2]$, $\sigma = 1.03$), and hot ($\mathcal{U}[1.4,2]$, $\sigma = 1.14$) states, running evolutionary dynamics for $4\times10^5$ replacements [Fig.~\ref{fig:branch}(d); $N = 256$, $24$ disorder realizations]. The hot, warm, and cold initializations all converge to close to criticality $\sigma \approx 0.95$,
 and $\langle R \rangle = 1.44$--$1.45$. The fit-free mean trait [inset] shows the convergence most cleanly: three ergodic starts reach $\langle T\rangle \approx 0.12$--$0.14$ within ${\sim}\,2\times 10^3$ steps.

The frozen start is the exception: it settles into a distinct, colder state ($\langle T\rangle \approx 0.05$, $\sigma \approx 0.85$) where cascades heal before stranding any established agent. This is a metastable branch --- the degenerate limit of~(a).

The edge remains the attractor for any ergodic or mildly cold start.

\emph{(d) Finite-size approach to criticality.} A self-organized branching process is exactly critical only as $N\to\infty$; finite-$N$ corrections, controlled by the single-site drive~\cite{zapperi1995self}, distinguish genuine SOC from a system parked near a crossover. The post-transient activity fit [Fig.~\ref{fig:branch}(c); $24$ realizations per size, $8$ at $N = 8192$] gives $\sigma$ increasing monotonically from $0.88$ at $N = 128$ and saturate at $\approx 1.01$ at $N \geq 2048$. Fits including the seed transient inflate large-$N$ values to $1.02$--$1.03$, a bias of the estimator (Appendix~\ref{sec:branch-est}).

The seed transient is itself informative and explains why it must be excluded from the branching-ratio fit. Flipping the seed agent~$j$ shifts each agent~$i$'s local field by an amount proportional to $|J_{ij}|$; from Eq.~\eqref{eq: chi-def}, agent~$i$ then flips with probability $\alpha\,\chi_i\,|J_{ij}|$. Averaging over the SK couplings, $\mathbb{E}[|J_{ij}|] = \sqrt{2/(\pi N)}$, and summing over all $N$ potential targets gives the first-step activity
\begin{equation}
\langle A(0)\rangle \;=\; \alpha\,\sqrt{2N/\pi}\,\;\langle\chi\rangle ,
\label{eq: seed-activity}
\end{equation}
This initial activity is set by the all-to-one coupling to the seed, not by the branching rate~$\sigma$ that governs subsequent steps; including it in the geometric fit therefore inflates the large-$N$ estimates of~$\sigma$ (the bias noted above). The equation also reveals the warm-agent depletion: an $N$-independent seed requires $\langle\chi\rangle \sim N^{-1/2}$, i.e.\ the soft fraction must scale as $n_{\mathrm{soft}}/N \sim N^{-1/2}$ --- a depletion supplied by the evolved trait distribution, not by any fixed uniform temperature. (Here \emph{soft} refers to the $\Theta(\sqrt{N})$ agents with susceptibility $\chi$ of order unity, $|Q| \lesssim T$, a subset of the $\Theta(N)$ \emph{warm} agents that carry the cascades and set $\sigma = aw$; most warm agents are deep in their wells and stiff, contributing to the branching ratio but not to the seed activity.)

\emph{(e) The avalanche exponent, and generality.} For a critical Galton--Watson process with finite offspring variance, $P(S) \sim S^{-3/2}$~\cite{pruessner2012self,zapperi1995self}. The exponent is \emph{derived} from $\sigma \to 1$, not assumed, and is universal: it depends only on marginal branching with finite variance. The measured $\tau \approx -1.58$ (Fig.~\ref{fig:epsart-2}) and $\sigma \to 1$ are two measurements of one fact. All three signatures --- ergodicity edge, mean-field avalanche exponent, and reward optimum --- collapse onto a single mechanism: extremal selection drives the population to the temperature at which its learning dynamics is marginally stable.

The analysis of subsection~(b), and its central result --- that the critical point is the stationary attractor in the thermodynamic limit --- apply more generally. Any system with (i) extremal replacement, (ii) a mutation kernel of full support, so that every class must sustain a death flux, and (iii) a hard class of agents whose mortality rate vanishes faster than any power of $N$ for subextensive events, will sit at critical collective dynamics in its stationary state, because subcriticality starves the hard class of stranding and supercriticality starves it of selection. This explains why the SOC phenomenon is robust to different inheritance rules (Fig.~\ref{fig:inherit}). A notable exception is when the system is initialized in the frozen state: it remains there and does not reach the critical point, at least on accessible timescales. This is a key difference from other well-known SOC models, such as the sandpile model, where the system is always driven to the critical point regardless of the initial state.

\subsection{\label{sec:theory-search}Why criticality confers optimality}\label{sec:theory-stab}

Having established why the system self-organizes to criticality, we now look at why this critical state is also optimal. We give the physical picture first, then make it quantitative through local stability, whose per-agent value \emph{is} the reward.

\emph{What is being optimized.} Summing the rewards of Eq.~\eqref{eq: SK-rwd},
\begin{equation}
\langle R \rangle \;=\; \frac{1}{N}\sum_i \sigma_i \sum_j J_{ij}\sigma_j \;=\; -\,\frac{2E}{N},
\quad E = -\!\!\sum_{i<j} J_{ij}\sigma_i\sigma_j ,
\label{eq: R-is-energy}
\end{equation}
``Collective optimality'' is the canonical hard problem of finding low-energy states of a rugged landscape, which is why learning alone terminates in satisficing states~\cite{garnier2024unlearnable}. The branching ratio describes how the population \emph{moves} on that landscape: subcritical is a greedy quench; supercritical is chaotic; at $\sigma = 1$ rearrangements span every scale, small ones refining the current basin and rare large ones carrying the population into another. Only the last is what a stochastic optimizer needs.

\emph{Why the drive must be extremal.} Perturbing the \emph{worst} variable is the principle behind extremal optimization~\cite{boettcher2002optimization}: it escapes greedy freezing while preserving the rest of the configuration. {\tt EvoSK} implements this automatically and, additionally, self-tunes to scale-free cascades. A null model replacing a \emph{random} agent instead of the extremal one abolishes self-organization ($\langle T \rangle = 1.007$) and collapses the reward to $\langle R \rangle = 0.849 \pm 0.008$ against $1.455 \pm 0.006$ for {\tt EvoSK}. Extremal selection, not replacement per se, does the key work.

The following three subsections make the picture quantitative.

\emph{(a) The reward is the local stability.} Define the local stability $\lambda_i(t) \equiv \sigma_i(t)\, h_i(t)$, where $h_i(t) = \sum_j J_{ij}\sigma_j(t)$ is the local field. Comparing with Eq.~\eqref{eq: SK-rwd} gives the exact identity $R_i(t) = \lambda_i(t)$: each agent's reward is its local stability. The collective reward is therefore the first moment of the stability distribution~$p(\lambda)$. Since fitness $f_i = \sigma_i Q_i$ and $Q_i$ is the exponentially weighted moving average of $h_i$ [Eq.~\eqref{eq: SK-q-update}], fitness is a smoothed copy of~$\lambda$ with the same sign structure (Pearson correlation $0.997$ at $\alpha = 0.75$). The least-fit agent is therefore, to excellent accuracy, the least locally stable, and {\tt EvoSK} is extremal dynamics on the stability field --- the rule ``repair the most unstable site''~\cite{boettcher2002optimization}. The cascades of Sec.~\ref{sec:theory-branch} are cascades of \emph{stability repair}.

\begin{figure*}[t]
\includegraphics[width=\linewidth]{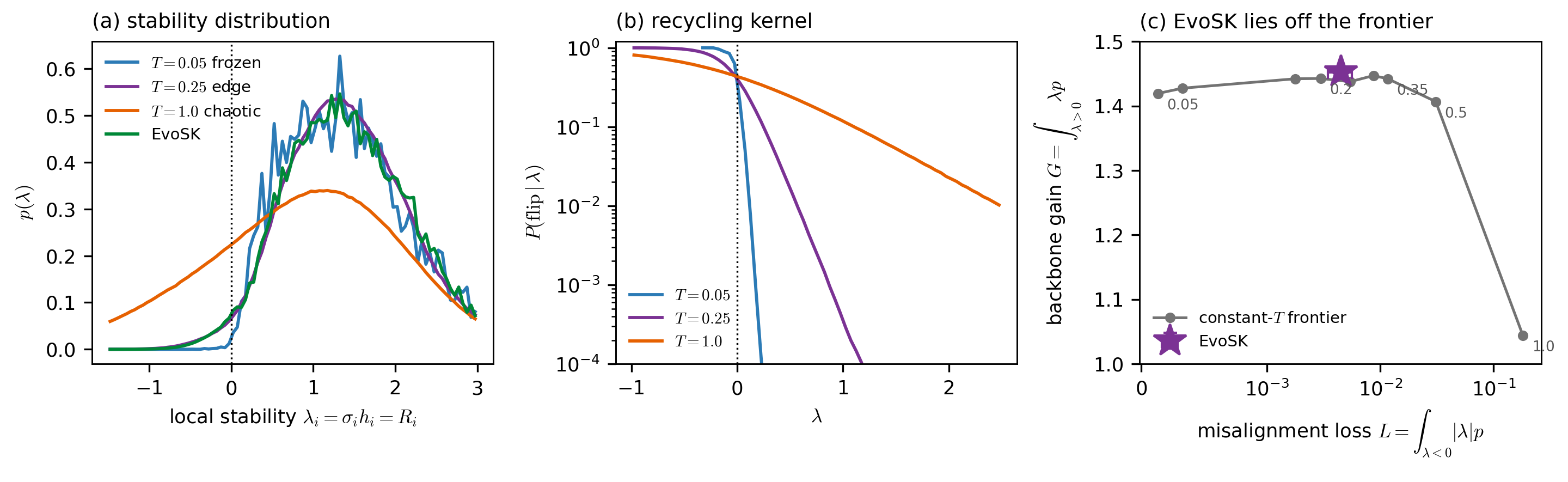}
\caption{\label{fig:stab}\textbf{Local-stability account of collective optimality} ($\alpha = 0.75$).
(a)~Distribution of the local stability $\lambda_i = \sigma_i h_i = R_i$, whose mean is the collective reward and, by Eq.~\eqref{eq: R-is-energy}, minus twice the energy density. The frozen population has almost no weight at $\lambda<0$; the chaotic one has $23\%$ of agents misaligned; the edge and the evolved state are intermediate.
(b)~The flip kernel $P(\mathrm{flip}\,|\,\lambda)$ is logistic in $\lambda$ with a width that grows with $T$: selective at the edge, indiscriminate in the chaotic phase.
(c)~The gain--loss plane of Eq.~\eqref{eq: GL}. Constant-temperature systems are confined to a one-parameter frontier (grey, labelled by $T$) because a single kernel width must trade backbone quality against misalignment. The evolved population (star) lies off the frontier: it attains the frontier's maximal backbone quality $G$ at roughly half the misalignment cost $L$ of the uniform temperature that matches it ($T = 0.30$). Error bars are the SE over eight paired disorder realizations; the excess in $G$ over the frontier maximum is within the statistical uncertainty, the reduction in $L$ is significant ($3.2$\,SE), and the panel should be read in the horizontal direction.}
\end{figure*}

\emph{(b) The flip kernel and the gain--loss trade-off.} The softmax policy gives a one-step flip probability logistic in stability, $P(\mathrm{flip}\,|\,\lambda) = [1+e^{2\lambda/T}]^{-1}$, of width $w = T/2$, confirmed by measurement [Fig.~\ref{fig:stab}(b)]. The kernel width~$w$ sets the trade-off between exploration and stability. Writing $\langle R \rangle = G - L$ with
\begin{equation}
G = \!\int_{\lambda>0}\!\! \lambda\, p(\lambda)\,d\lambda, \quad
L = \!\int_{\lambda<0}\!\! |\lambda|\, p(\lambda)\,d\lambda ,
\label{eq: GL}
\end{equation}
$G$ measures backbone quality (contribution from aligned agents) and $L$ the cost of misalignment (contribution from unstable ones). Raising $T$ widens the kernel: the backbone deepens ($G$ rises) because a wider kernel explores more of the landscape, but misalignment grows ($L$ rises too) because more agents are destabilized. A cold population pays no $L$ yet cannot improve $G$ because its narrow kernel cannot explore. For a homogeneous population these two channels are locked by the single parameter~$w(T)$, tracing a one-parameter frontier in $(L,G)$ [Fig.~\ref{fig:stab}(c)], with the optimum at the edge where backbone improvement first begins to outpace misalignment growth.

\emph{(c) Why evolution beats every fixed temperature.} The frontier of Fig.~\ref{fig:stab}(c) is a constraint of homogeneity: a uniform-temperature population must use one kernel width for both exploration and stability, so it cannot improve $G$ without paying~$L$. The evolved trait distribution breaks this constraint. The cold majority supplies backbone at essentially zero $L$ (their narrow kernels keep them aligned), while the hot minority --- continually resupplied by mutation and culled by selection before it accumulates misalignment --- provides the wide-kernel exploration that deepens $G$. The measured evolved values are $G = 1.453 \pm 0.008$ and $L = 0.0045 \pm 0.0011$. In backbone quality the evolved state matches the frontier maximum ($G = 1.447 \pm 0.007$ at $T = 0.30$; paired excess $+0.006 \pm 0.009$). The gain is in misalignment: the evolved population pays only $52\%$ of the frontier's $L$ at that temperature ($\Delta L = -0.0042 \pm 0.0013$, $3.2$\,SE).

This is the mechanism underlying Fig.~\ref{fig:opt}(b).

\bibliography{references}

\end{document}